\documentclass[aps,prb,reprint,amsmath,amssymb,longbibliography,superscriptaddress,floatfix]{revtex4-2}
\usepackage{graphicx}
\usepackage{amsmath}
\usepackage{amssymb}
\usepackage{mathrsfs}
\usepackage{bm}
\usepackage{xcolor}
\usepackage{hyperref}
\hypersetup{
  hidelinks,
  pdftitle={Hydrodynamic Anomalous Hall Effect in a Two-Dimensional Dirac Fluid},
  pdfauthor={A. Levchenko and D. A. Pesin}
}
\usepackage{pgfplots}
\pgfplotsset{compat=1.16}
\newcommand{\zhat}{\hat{\bm z}}
\newcommand{\kk}{\bm k}
\newcommand{\pp}{\bm p}
\newcommand{\qq}{\bm q}
\newcommand{\rr}{\bm r}
\newcommand{\EE}{\bm E}
\newcommand{\vv}{\bm v}
\newcommand{\eF}{\varepsilon_F}
\newcommand{\dr}{\delta\rr}
\newcommand{\nB}{n_{\rm B}}
\newcommand{\hc}{\mathcal H}

\begin{document}

\title{Hydrodynamic Anomalous Hall Effect in a Two-Dimensional Dirac Fluid}

\author{A. Levchenko}
\affiliation{Department of Physics, University of Wisconsin--Madison, Madison, Wisconsin 53706, USA}

\author{D. A. Pesin}
\affiliation{Department of Physics, University of Virginia, Charlottesville, VA 22904, USA}

\begin{abstract}
In ultraclean conductors, electron-electron collisions can make transport hydrodynamic.
In a Berry-curved band, these collisions can also produce two-particle coordinate shifts, which contribute to
the anomalous Hall current in the presence of a transport electric field. We develop a kinetic theory of this contribution for a
two-dimensional massive Dirac Fermi liquid with weak contact interactions and weak
short-range disorder. The computed response is independent of interaction strength at leading Born order. Its contribution to the anomalous Hall conductivity is comparable to the thermal correction to the intrinsic anomalous Hall conductivity. The existence of the momentum zero mode of the electron-electron collision integral leads to a nontrivial frequency dependence of the anomalous Hall response considered in this work. In particular, the limits of vanishing frequency (dc) and vanishing disorder do not commute. In the dc limit, the momentum response to the transport field grows inversely with the momentum relaxation rate. This allows arbitrarily weak, energy-dependent impurity scattering to produce a finite correction by coupling conserved momentum to the modes relaxed by electron-electron scattering. This correction disappears when the clean limit is taken before the zero-frequency limit. As a result, the interaction-induced Hall response retains information about the mechanism of momentum relaxation even as the overall disorder strength tends to zero.
\end{abstract}

\maketitle

\section{Introduction}\label{sec:intro}

The anomalous Hall effect (AHE), a transverse voltage in response to a longitudinal current in
the absence of an external magnetic field, has accompanied condensed matter physics for more
than a century, and its theory for seven decades. Karplus and Luttinger discovered that Bloch
electrons in a spin-orbit-coupled band acquire an anomalous velocity transverse to an applied
electric field, producing a dissipationless Hall current entirely of band-structure origin
\cite{karplus54}. Disorder provides additional sources of Hall current:
Smit identified the contribution from asymmetric scattering probabilities, or skew scattering
\cite{smit55,smit58}. Berger identified the sideways displacement of a wave packet in a single scattering event,
the side jump \cite{berger70}; the ensuing
controversy over the relative roles of the intrinsic and extrinsic mechanisms lasted for decades
(see also Ref.~\cite{luttinger58}). The modern synthesis rests on the semiclassical wave-packet
theory of Bloch dynamics \cite{sundaram99}: the Karplus-Luttinger anomalous velocity is the
Berry curvature of the band acting as a magnetic field in momentum space
\cite{jungwirth02,xiao10}, the intrinsic Hall conductivity is the Fermi-sea integral of that
curvature, reducible by an argument of Haldane to a Fermi-surface property
\cite{haldane04}, and its quantized limit in a filled band is the Chern number of
Thouless-Kohmoto-Nightingale-den~Nijs \cite{tknn82}. Haldane's model established
that a quantized Hall state can exist without a net magnetic flux \cite{haldane88};
the quantized anomalous Hall effect was later realized experimentally in magnetically doped
topological-insulator films \cite{chang13}.
Comprehensive accounts of this development may be found in the reviews
\cite{nagaosa10,sinitsyn08,xiao10} and, in a form close to the perspective of the present work,
in Ref.~\cite{konig21}.

The sobering counterpoint to this conceptual richness is that nearly all of it
(the semiclassical kinetics, the Kubo-St\v{r}eda diagrammatics, and the delicate cancellations
between intrinsic and disorder-induced contributions
\cite{sinitsyn08,sinitsyn07,ado15}) addresses a problem posed at the single-particle level:
one electron in a chiral band, scattering off static disorder. The role of electron--electron
interactions in the AHE remains comparatively unexplored, and the results that do exist tie the
interaction corrections to impurities in an essential way. Langenfeld and W\"olfle showed that
the Altshuler-Aronov-type quantum correction to the anomalous Hall conductivity vanishes
identically \cite{langenfeld91}, a statement later extended to arbitrary strength of the impurity
scattering on general symmetry grounds by Muttalib and W\"olfle \cite{muttalib07}, so that the
familiar diffusive interaction corrections of longitudinal transport have no counterpart in
$\sigma_{xy}$. It was recently demonstrated that the Cooper channel is an exception: interaction
corrections from superconducting fluctuations, relevant even in nominally non-superconducting
materials, generate a temperature-dependent correction to the extrinsic Hall response. It behaves asymptotically as $\sqrt{T}/\ln(T_0/T)$ in three dimensions and as $\ln\ln(T_0/T)$ in
two \cite{li20}, where $T$ is the temperature. Both classes of results describe interference between electron--electron and
electron-impurity scattering: remove the impurities, and the corrections disappear together with
the extrinsic AHE they correct. Interaction corrections that do not require disorder have been
studied perturbatively: in the massive Dirac model the complete first-order correction contains
self-energy and vertex contributions that combine to produce a
temperature-dependent correction to the intrinsic Hall conductivity \cite{dumitriu24}; such
first-order corrections modify the single-particle response rather than arising from a collision
integral. The question of what an interacting chiral liquid does on its own through collisions remained open until very
recently. Can electron-electron collisions, which conserve momentum and are usually deemed
invisible in the conductivity, generate a Hall response of a pristine system? That question has now acquired both experimental urgency and a theoretical answer in principle.

On the experimental side, ultraclean conductors, graphene above all, have entered the
hydrodynamic regime, in which electron-electron collisions dominate over both disorder and
phonon scattering and electrons flow as a viscous fluid. The Dirac fluid near charge neutrality
has been observed through the dramatic breakdown of the Wiedemann-Franz law \cite{crossno16};
viscous whirlpools and negative nonlocal resistance have been observed \cite{bandurin16}; and
a transverse hydrodynamic response, the Hall viscosity, has been
measured in graphene channels \cite{berdyugin19}. Of particular relevance, transverse topological
currents associated with Berry curvature in gapped graphene superlattices have been probed
through nonlocal valley signals \cite{gorbachev14}: band chirality and hydrodynamic flow are therefore accessible in related graphene platforms. 

On the theory side, the hydrodynamic transport coefficients of the Dirac fluid are
comprehensively developed, from the quantum-critical collision-limited intrinsic conductivity
\cite{fritz08,kashuba08} to the nearly perfect fluidity encoded in the shear viscosity
\cite{muller09}; broader accounts appear in Refs.~\cite{narozhny19,lucas18}.
The conductivity of charge-neutral graphene also receives a contribution from hydrodynamic
fluctuations \cite{levchenko2026}. 

A new paradigm for the Hall response emerged with the
realization that the AHE survives in a pristine,
interaction-dominated chiral Fermi liquid~\cite{pesin18}. That work generalized the two-particle collisional shifts, introduced in Ref.~\cite{son2014} for Lorentz-invariant systems, to the solid-state context, and showed that these shifts drive a Hall current in the presence of a transport electric field. 

The present work is motivated by these developments, and its purpose is to provide a quantitative theory
of the coordinate-shift contribution in a concrete model. We choose the paradigmatic model of two-dimensional massive Dirac fermions: the minimal band
structure with all required ingredients (broken time reversal, Berry curvature,
nonparabolicity), and historically the most thoroughly studied host of the AHE. The intrinsic
contribution and its half-quantized limit were established early \cite{ludwig94,sinitsyn07},
and the complete single-particle answer with disorder, whose Hall conductivity is strongly
suppressed at high Fermi energies, was obtained in Ref.~\cite{ado15}. The conductivity and viscosity results discussed above concern the closely related
massless Dirac fluid near charge neutrality. The present calculation extends this kinetic description
to the collisional Hall response of a degenerate, massive band.

Our central results are as follows. We show that the collisional drive due to the two-particle coordinate shift is exactly orthogonal to all zero modes of the electron-electron collision operator. A finite-frequency kinetic treatment then yields
the interparticle Hall response in both the clean limit, defined as 
$\Gamma\ll\omega\ll\gamma_{\rm ee}$, and the dc limit, in which
$\omega\ll\Gamma\ll\gamma_{\rm ee}$, with weak scalar Gaussian white-noise disorder
providing momentum relaxation. Here $\Gamma$ is the momentum-relaxation rate and
$\gamma_{\rm ee}$ characterizes relaxation of the relevant nonconserved modes by
electron-electron collisions.

In the clean limit, the Hall response is carried by distribution perturbations with zero net momentum. It therefore vanishes for a parabolic band, where electric current is proportional to total momentum. The clean limit Hall conductivity is given by
\[
\sigma^{\rm cl}_{xy}\simeq
\frac{e^2}{\hbar}\,t^2\mathcal H(r;t),
\qquad
r=\frac{m}{\varepsilon_F},\quad
t=\frac{T}{\varepsilon_F},
\]
per Dirac cone. To illustrate its magnitude and mass dependence, we evaluate the dimensionless coefficient at a representative low temperature \(t=0.02\). Numerical inversion of the collision operator gives
\[
\mathcal H(r;0.02)\simeq0.29\,r,
\qquad 0.1\leq|r|\leq0.8,
\]
\textit{i.e.} a nearly linear mass dependence away from the band edge. The temperature argument of $\mathcal H(r;t)$ has to be retained because thermally regulated collinear scattering can produce residual temperature dependence beyond the explicit \(T^2\) prefactor; this dependence is examined in Sec.~\ref{sec:logs}.

In the dc limit, the same generation vector and
collision matrix enter, but the effective drive retains the energy dependence of impurity
scattering. To leading thermal order,
\begin{equation}
\sigma^{\rm dc}_{xy}\simeq\frac{2}{1+3r^2}\,\sigma^{\rm cl}_{xy}.
\end{equation}
We also evaluate both coefficients with the full thermal width retained. The disorder
strength cancels from the leading dc result, while the coupling of conserved momentum to
relaxing modes leaves a finite correction. The limits of vanishing frequency and vanishing
disorder therefore do not commute. Both shift accumulation and anomalous distribution acquire
the correction and remain equal.

Both limiting contributions are independent of interaction strength at leading Born order
within the hydrodynamic hierarchy and carry an explicit $T^2$ prefactor. In the convention
$j_i=\sigma_{ij}E_j$, the intrinsic conductivity of the same cone is
$\sigma^{\rm int}_{xy}=-(e^2/4\pi\hbar)(m/\eF)$ at zero temperature, so the computed
interparticle term opposes it. At the two representative masses sampled, the clean
leading-Sommerfeld coefficient changes by less than eight percent when temperature is
lowered by a factor of forty; the asymptotic ratio of the collinearly enhanced collision
moments remains unresolved. We compare the two limiting shift responses with the intrinsic
Sommerfeld correction. Impurity side-jump and skew contributions must themselves be
recalculated in the interacting fluid; their hydrodynamic modification and the finite-channel
boundary problem are beyond the present calculation.

The rest of the paper is organized as follows. Section~\ref{sec:shifts} introduces the model, disorder, coordinate shifts, and kinetic equation.
Section~\ref{sec:conservation} establishes the conservation laws and eliminates the momentum
mode at arbitrary $\omega/\Gamma$. Section~\ref{sec:conductivity} evaluates the two limiting
Hall coefficients and their frequency crossover. Section~\ref{sec:summary} compares them
with the intrinsic background and outlines open transport problems. The massive Dirac shift
is derived in Appendix~\ref{sec:dirac}; the remaining technical derivations, including the
exact momentum-mode elimination, are collected in
Appendixes~\ref{app:sj}--\ref{app:disorder-projection}.

\begin{figure}[!htbp]
\includegraphics[width=\columnwidth]{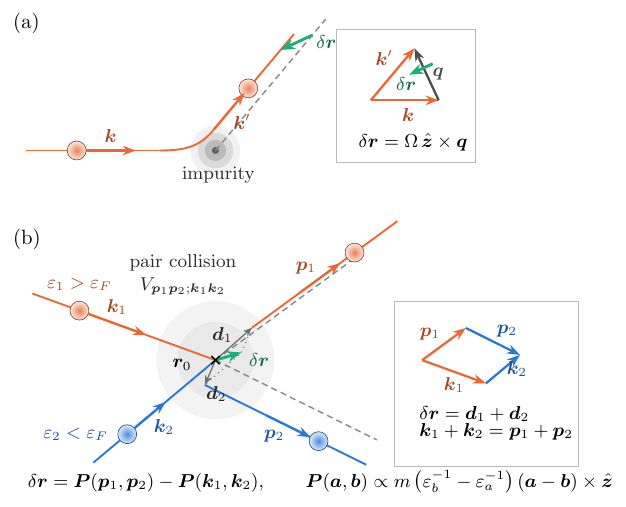}
\caption{Coordinate shifts in single-particle and two-particle scattering. Displacements are
exaggerated; dashed lines are reference outgoing trajectories with the same final momenta but
without a coordinate shift. (a)~An impurity collision changes the wave-packet momentum from
$\kk$ to $\kk'$. At small transfer $\qq=\kk'-\kk$, the leading shift is
$\dr=\Omega(\kk)\zhat\times\qq$; the inset illustrates its orientation for $\Omega>0$.
For the conduction band studied here, $\Omega<0$ when $m>0$.
(b)~A pair collision conserves $\kk_1+\kk_2=\pp_1+\pp_2$. The offsets $\bm d_1,\bm d_2$
illustrate the total shift $\dr=\bm d_1+\bm d_2$ of Eq.~\eqref{eq:master}; they do not assign
separate shifts to identifiable electrons. The pair center of charge shifts by $\dr/2$, so the
transported charge displacement is $e\dr$ and the field does work $e\EE\cdot\dr$.
For the contact Dirac model, $\dr=\bm P(\pp_1,\pp_2)-\bm P(\kk_1,\kk_2)$, which vanishes
when all four states lie on a common energy shell. The colors show one possible thermally
active configuration, not a requirement that the incoming energies straddle $\eF$.
The trajectory shapes and momentum arrows are schematic.}
\label{fig:shift}
\end{figure}

\section{Collisional coordinate shifts and the kinetic equation}
\label{sec:shifts}

In this section, we formulate a Boltzmann description of the hydrodynamic anomalous Hall effect, incorporating the two-particle coordinate shifts of Ref.~\cite{pesin18} into the semiclassical kinetic equation. We calculate the interparticle coordinate-shift contribution to the Hall response; other anomalous Hall mechanisms are outside the scope of this calculation.

We consider a spatially uniform situation, excluding momentum relaxation at channel walls. Since the electron-electron collisions considered here conserve quasimomentum, a stationary response to a dc electric field requires an additional momentum sink. We take this to be weak impurity scattering; momentum transfer to phonons provides another possibility. The hydrodynamic regime is defined by
$\{\Gamma,\omega\}\ll \gamma_{\rm ee}$, 
where $\Gamma$ is the momentum-relaxation rate due to impurities, $\omega$ is the driving frequency, and $\gamma_{\rm ee}$ characterizes relaxation of the relevant nonconserved modes by electron--electron collisions.

Within the hydrodynamic regime, \(\Gamma,\omega\ll\gamma_{\rm ee}\), we distinguish the clean limit, \(\Gamma\ll\omega\), from the dc limit, \(\omega\ll\Gamma\). In the clean limit, momentum relaxation is taken to zero before the driving frequency; in the dc limit, the frequency is taken to zero at finite momentum relaxation, after which the disorder strength may be made arbitrarily small. These two orders of limits generally yield different interparticle Hall conductivities.

\subsection{Model, disorder, and frequency regime}
\label{sec:disorder-model}

We consider a single nondegenerate massive Dirac cone,
\begin{equation}
H_0=v(k_x\sigma_x+k_y\sigma_y)+m\sigma_z,\qquad
\varepsilon_{\kk}=\sqrt{v^2k^2+m^2},
\label{eq:hamiltonian}
\end{equation}
with a weak contact repulsion $U=\lambda$.
We set $\hbar=k_{\rm B}=1$ in intermediate expressions, hold $\mu=\eF>|m|$ fixed unless stated otherwise,
and define $p_F=\sqrt{\eF^2-m^2}/v$, $v_F=v^2p_F/\eF$, and $\nu_F=\eF/(2\pi v^2)$.
The analytic thermal expansion assumes $T\ll\eF-|m|$, and the Born collision kernel assumes
$\nu_F\lambda\ll1$. Appendix~\ref{sec:dirac} gives the spinors and the complete pair-shift derivation.

Momentum relaxation is supplied by a weak scalar Gaussian potential,
$H=H_0+U_{\rm dis}(\rr)$, with
\begin{equation}
\overline{U_{\rm dis}(\rr)}=0,\qquad
\overline{U_{\rm dis}(\rr)U_{\rm dis}(\rr')}
=\mathcal W\delta^{(2)}(\rr-\rr').
\label{eq:dis-correlator}
\end{equation}
For elastic scattering through an angle $\alpha$ the conduction-band overlap is
\begin{equation}
|\langle u_{\kk'}|u_{\kk}\rangle|^2
=\frac{1+m^2/\varepsilon^2+(1-m^2/\varepsilon^2)\cos\alpha}{2}.
\label{eq:dis-overlap}
\end{equation}
On either first angular harmonic, $F(\varepsilon)\cos\vartheta$ or
$F(\varepsilon)\sin\vartheta$, the symmetric Born impurity collision operator
$\hat D$ is multiplication by the transport rate
\begin{equation}
\begin{aligned}
\gamma_{\rm tr}(\varepsilon)
&=2\pi\mathcal W\nu(\varepsilon)
\int\frac{d\alpha}{2\pi}|\langle u_{\kk'}|u_{\kk}\rangle|^2(1-\cos\alpha)\\
&=\kappa\frac{\varepsilon^2+3m^2}{\varepsilon},
\qquad \kappa=\frac{\mathcal W}{4v^2},
\end{aligned}
\label{eq:dis-rate}
\end{equation}
where $\nu(\varepsilon)=\varepsilon/(2\pi v^2)$. This agrees with the white-noise limit of
Ref.~\cite{ado17}. The energy dependence of this transport rate is essential; replacing it by
its Fermi-surface value before solving the momentum balance changes the dc answer.

We use homogeneous fields proportional to $e^{-i\omega t}$ and retain arbitrary
$\omega/\Gamma$ within the hydrodynamic regime specified above. Here $\Gamma$ is the
thermally weighted momentum-relaxation rate defined in Eq.~\eqref{eq:momentum-moments};
$\gamma_{\rm ee}$ denotes relaxation of the relevant nonconserved first-harmonic modes, rather
than the quasiparticle decay rate. The collision broadening is small compared with $T$.
We use additive electron-electron and impurity collision operators and neglect disorder
renormalization of the band and of the interaction vertex.

The calculated Hall term is linear in the interparticle coordinate shift. The symmetric
impurity operator is retained in the distribution, but impurity coordinate-shift and skew
vertices and the intrinsic Berry current are separate contributions. Gaussian disorder does not
remove skew scattering on impurity pairs \cite{ado17}. Thus the result below is the interparticle
shift contribution, not the total Hall conductivity of the disordered metal.

\subsection{Side jump in single-particle scattering}
\label{sec:sj}

The elementary object of the theory is the coordinate shift (side jump) accumulated by a Bloch
wave packet in a single scattering event \cite{sinitsyn08,sinitsyn07,xiao10}. For a packet built
from the states of band $n$ that scatters off a static perturbation with matrix element
$V_{\kk'\kk} = \langle u_{\kk'}|\hat V|u_{\kk}\rangle$, a stationary-phase analysis of the
incoming and outgoing packet centers (Appendix~\ref{app:sj}) gives
\begin{equation}
\dr_{\kk'\kk} \;=\; \bm A(\kk') - \bm A(\kk)
- \big(\nabla_{\kk'}+\nabla_{\kk}\big)\arg V_{\kk'\kk},
\label{eq:sj}
\end{equation}
where
\begin{equation}
\bm A(\kk) = i\langle u_{\kk}|\nabla_{\kk} u_{\kk}\rangle
\end{equation}
is the intraband Berry connection. Equation \eqref{eq:sj} is invariant under the band-gauge
transformation $|u_{\kk}\rangle \to e^{i\beta(\kk)}|u_{\kk}\rangle$: the connections shift by
$-\nabla\beta(\kk')$ and $+\nabla\beta(\kk)$, while
$\arg V_{\kk'\kk} \to \arg V_{\kk'\kk} - \beta(\kk') + \beta(\kk)$, and the gradient term
compensates exactly. It is antisymmetric under $\kk\leftrightarrow\kk'$ because
$V_{\kk\kk'} = V^*_{\kk'\kk}$ for a Hermitian perturbation.

For a scalar scatterer the matrix element factorizes as
$V_{\kk'\kk}=U_{\rm s}(\kk'-\kk)\langle u_{\kk'}|u_{\kk}\rangle$,
where $U_{\rm s}$ is the Fourier amplitude of the scattering potential. Its phase makes
no contribution to the shift because
\[
(\nabla_{\kk'}+\nabla_{\kk})\arg U_{\rm s}(\kk'-\kk)=0.
\]
Thus, for small momentum transfer $\qq=\kk'-\kk$, Eq.~\eqref{eq:sj} reduces to
a purely geometric expression (Appendix~\ref{app:sj}),
\begin{equation}
\dr = \Omega(\kk)\,\zhat\times\qq + O(q^2),
\qquad \Omega = \big[\nabla\times\bm A\big]_z .
\label{eq:sjsmallq}
\end{equation}
The gauge-dependent gradient pieces cancel and only the Berry curvature survives: the small-angle
side jump is a rotation of the momentum transfer by $\pi/2$, weighted by $\Omega$. This is the
elementary building block of the extrinsic AHE mechanisms reviewed in Ref.~\cite{sinitsyn08}.

\subsection{Two-particle generalization}
\label{sec:2p}

The generalization to a pair collision \cite{pesin18} proceeds along the same lines, with the
single-particle matrix element replaced by the antisymmetrized two-body one. For the process
$(\kk_1,\kk_2)\to(\pp_1,\pp_2)$, with $\pp_1+\pp_2 = \kk_1+\kk_2$,
\begin{equation}
\begin{aligned}
V_{\pp_1\pp_2;\kk_1\kk_2}
&= U(\pp_1{-}\kk_1)\,\langle u_{\pp_1}|u_{\kk_1}\rangle\langle u_{\pp_2}|u_{\kk_2}\rangle\\
&\ - U(\pp_1{-}\kk_2)\,\langle u_{\pp_1}|u_{\kk_2}\rangle\langle u_{\pp_2}|u_{\kk_1}\rangle ,
\end{aligned}
\label{eq:V2}
\end{equation}
where $U(\qq)$ is the interaction form factor. Constructing two-packet in and out states and
carrying through the stationary-phase analysis, each packet contributes its Berry connection, and
the entire phase of the pair amplitude contributes through $\arg V$. The result is the
gauge-invariant two-particle coordinate shift
\begin{equation}
\begin{aligned}
\dr^{\pp_1\pp_2}_{\kk_1\kk_2}
&= \bm A(\pp_1)+\bm A(\pp_2)-\bm A(\kk_1)-\bm A(\kk_2)\\
&\quad- \big(\nabla_{\pp_1}+\nabla_{\pp_2}+\nabla_{\kk_1}+\nabla_{\kk_2}\big)
\arg V_{\pp_1\pp_2;\kk_1\kk_2}
\end{aligned}.
\label{eq:master}
\end{equation}
Gauge invariance follows exactly as for Eq.~\eqref{eq:sj}: under
$|u_{\kk}\rangle\to e^{i\beta(\kk)}|u_{\kk}\rangle$ the matrix element acquires the phase
$-\beta(\pp_1)-\beta(\pp_2)+\beta(\kk_1)+\beta(\kk_2)$, whose four-gradient cancels the shift of
the connections. Two structural properties follow directly: (i) symmetry under the simultaneous
relabeling $(1\leftrightarrow2)$ of both pairs, since $V$ changes only by an overall sign under
exchange of either pair; (ii) antisymmetry under interchange of initial and final states,
$\dr^{\kk_1\kk_2}_{\pp_1\pp_2} = -\dr^{\pp_1\pp_2}_{\kk_1\kk_2}$, from the Hermiticity of $\hat V$.
Physically, $\dr$ is the displacement of the pair's outgoing wave packets relative to the incoming
ones, accumulated during the collision; it is the interaction analog of the side jump, and, like
it, is entirely a property of the Bloch geometry and of the interaction phase, as illustrated in Fig.~\ref{fig:shift}.

\subsection{Kinetic equation with collisional shifts}
\label{sec:kinetic}

We retain the Hall contribution linear in the coordinate shift. Throughout the collision
formulas, $\sum_{\kk}\equiv\int d^2k/(2\pi)^2$, and $V$ in Eq.~\eqref{eq:V2} is the reduced
interaction amplitude with the momentum-conserving delta function removed. The shift enters
transport twice \cite{pesin18}. First, every collision displaces charge, so the
current acquires a shift contribution
\begin{equation}
\begin{aligned}
&\bm j^{\rm shift} = \frac{e}{4}\sum_{1234} W^{34}_{12}\,\dr^{34}_{12}\,
f_1 f_2 (1-f_3)(1-f_4),\\
&W^{34}_{12} = 2\pi\,|V_{34;12}|^2\,
(2\pi)^2\delta^{(2)}(\kk_1+\kk_2-\kk_3-\kk_4)\\
&\hspace{2cm}\times\delta(\varepsilon_1{+}\varepsilon_2{-}\varepsilon_3{-}\varepsilon_4),
\end{aligned}
\label{eq:jshift-def}
\end{equation}
where the integrals run over all ordered quadruples of momenta, and the factor $\tfrac14$ removes the
double counting of the initial and of the final pair of identical particles implied by the
antisymmetrized amplitude \eqref{eq:V2}. The current vanishes in equilibrium: relabeling
$(12)\leftrightarrow(34)$ leaves $W$ and, by energy conservation, the occupation product
invariant, while $\dr\to-\dr$. Throughout, $e$ denotes the (signed) carrier charge, so that
$e\EE$ is the force and $\bm j = e\sum_{\pp}\vv_{\pp}f_{\pp}$ is the ballistic
current. Second, in the presence of a
driving field the work $e\EE\cdot\dr$ done during the displaced collision modifies the energy
balance, $\delta(\varepsilon_1+\varepsilon_2-\varepsilon_3-\varepsilon_4)\to
\delta(\varepsilon_1+\varepsilon_2-\varepsilon_3-\varepsilon_4+e\EE\cdot\dr)$; expanding to linear
order in $\EE$ and using the equilibrium form of the occupations converts the collision integral
into a generation term. Both effects are captured by the linearized kinetic equation
\begin{equation}
\begin{aligned}
&\Big(-\frac{\partial f_0}{\partial\varepsilon}\Big)\partial_t\varphi_{\pp}
+e\EE\cdot\Big(\vv_{\pp}\,\frac{\partial f_0}{\partial\varepsilon} + \bm G_{\pp}\Big)
\\
&\qquad= I_{\rm ee}[\varphi_{\pp}]+I_{\rm dis}[\varphi_{\pp}],\\
&f_{\pp} = f_0 - \varphi_{\pp}\,\frac{\partial f_0}{\partial\varepsilon},
\end{aligned}
\label{eq:BE}
\end{equation}
where $I_{\rm dis}[\varphi]=-(-\partial_\varepsilon f_0)\hat D\varphi$ is the symmetric
impurity collision term. The interparticle generation vector is unchanged at the retained order:
\begin{equation}
\bm G_{\pp_1} = \frac{1}{2T}\sum_{234} W^{34}_{12}\,\dr^{34}_{12}\,
f_1 f_2(1-f_3)(1-f_4)\Big|_{f=f_0},
\label{eq:G}
\end{equation}
in which the factor $\tfrac12$ is the final-pair counterpart of the $\tfrac14$ in
Eq.~\eqref{eq:jshift-def}, the collision integral of particle 1 being
$I_{\rm ee}[f]_1 = \tfrac12\sum_{234}W^{34}_{12}[f_3f_4(1-f_1)(1-f_2)-f_1f_2(1-f_3)(1-f_4)]$.
Linearizing Eq.~\eqref{eq:jshift-def} in $\varphi$ and symmetrizing over the four legs [using the
$(1\leftrightarrow2)$, $(3\leftrightarrow4)$, $(12)\leftrightarrow(34)$ relabelings and
$-\partial_\varepsilon f_0 = f_0(1-f_0)/T$] expresses the shift current through the same vector:
with the decomposition $\varphi = \varphi^{v}+\varphi^{g}$ of the solution of Eq.~\eqref{eq:BE} into the
parts driven by the two source terms,
\begin{equation}
\bm j^{\rm shift} = e\sum_{\pp} \bm G_{\pp}\,\varphi^{v}_{\pp},
\qquad
\bm j^{\rm ball} = e\sum_{\pp} \vv_{\pp}\,
\Big({-}\frac{\partial f_0}{\partial\varepsilon}\Big)\varphi^{g}_{\pp} .
\label{eq:twocurrents}
\end{equation}
It is convenient to introduce the positive-semidefinite collision operator and inner product
\begin{equation}
\big(\hat C\varphi\big)_{\pp}\equiv -\,\frac{I_{\rm ee}[\varphi]_{\pp}}{-\partial_\varepsilon f_0},\quad
\langle\psi|\varphi\rangle=\int\!\frac{d^2p}{(2\pi)^2}\,\psi_{\pp}\varphi_{\pp}
\Big({-}\frac{\partial f_0}{\partial\varepsilon}\Big),
\label{eq:metric}
\end{equation}
i.e., $I_{\rm ee}[\varphi]_{\pp} = -(-\partial_\varepsilon f_0)(\hat C\varphi)_{\pp}$. In the
$\hbar=1$ convention used in the collision formulas, $\hat C$ has dimensions of energy
(equivalently inverse time); after restoring units, its physical rate operator is $\hat C/\hbar$.
A ket such as $|\varphi\rangle$ represents the whole momentum-dependent function
$\varphi(\pp)$, rather than a state at a particular momentum. An overlap integrates
over momentum and is a scalar; it has no remaining free momentum argument.
Self-adjointness of $\hat C$ in this metric follows from the four-fold
symmetrization of the linearized collision integral,
\begin{equation}
\langle\psi|\hat C|\varphi\rangle
=\frac{1}{8T}\sum_{1234}W^{34}_{12}\,f_1f_2(1-f_3)(1-f_4)
\Delta\psi\,\Delta\varphi,
\label{eq:entropy}
\end{equation}
where $\Delta\psi=\psi_1+\psi_2-\psi_3-\psi_4$, and similarly for $\Delta\varphi$.
For real trial functions this entropy-production form is symmetric and nonnegative at
$\psi=\varphi$. Its null space is spanned by $\{1,\pp,\varepsilon_{\pp}\}$.
At finite frequency we extend the real weighted pairing in Eq.~\eqref{eq:metric}
bilinearly to complex amplitudes, without complex conjugation. Reciprocity then means
transpose symmetry, not Hermitian self-adjointness at fixed $\omega$.

Defining the normalized generation function $\bm g(\pp)\equiv\bm G_{\pp}/(-\partial_\varepsilon f_0)$, the kinetic equation becomes
\begin{equation}
\begin{aligned}
\hat L_\omega\varphi&=e\EE\cdot(\vv-\bm g),\\
\hat L_\omega&=\hat C+\hat D-i\omega,\qquad
\hat R_\omega=\hat L_\omega^{-1}.
\end{aligned}
\label{eq:finite-BE}
\end{equation}
Thus $\varphi^v=e\hat R_\omega\EE\cdot\vv$ and
$\varphi^g=-e\hat R_\omega\EE\cdot\bm g$. The two currents in
Eq.~\eqref{eq:twocurrents} yield, with $j_i=\sigma_{ij}E_j$,
\begin{equation}
\sigma_{xy}^{\rm shift}(\omega)=e^2\Big[
\langle g_x|\hat R_\omega|v_y\rangle
-\langle v_x|\hat R_\omega|g_y\rangle\Big].
\label{eq:sigma-formal}
\end{equation}
In the isotropic model, $\bm g=g_\perp\zhat\times\hat\pp$
(Sec.~\ref{sec:vector}), so $g_x=-g_\perp\sin\vartheta$ and
$g_y=g_\perp\cos\vartheta$. The sine and cosine sectors have the same radial
resolvent, and both collision operators are symmetric. Consequently
\begin{equation}
\begin{aligned}
\sigma_{xy}^{\rm acc}(\omega)&=\sigma_{xy}^{\rm ad}(\omega)
=-e^2\langle v_x|\hat R_\omega|g_y\rangle,\\
\sigma_{xy}^{\rm shift}(\omega)&=-2e^2\langle v_x|\hat R_\omega|g_y\rangle.
\end{aligned}
\label{eq:sigma-2x}
\end{equation}
The labels ``acc'' and ``ad'' denote shift accumulation and anomalous distribution,
respectively. Each term is transverse, and their equality holds before either frequency limit
is taken. The remaining task is to invert the common
kinetic operator without discarding its weakly relaxed momentum mode.

\section{Conservation laws and the structure of the collisional drive}
\label{sec:conservation}

The contact shift vanishes on the common energy shell [Appendix~\ref{sec:dirac},
Eq.~\eqref{eq:P}]; its thermally active part is linear in the departure from that shell.
We now combine this kinematic property with the exact conservation laws and then eliminate
the weakly relaxed momentum mode at arbitrary $\omega/\Gamma$.

The difference between the dc and clean limits introduced in Sec.~\ref{sec:shifts} originates in the momentum zero mode of the electron-electron collision operator. An electric field excites both this conserved mode and modes relaxed by electron-electron collisions. To leading order in the hydrodynamic regime, the driven momentum response is proportional to $(\Gamma-i\omega)^{-1}$. For $\omega\gg\Gamma$, the oscillating field produces a bounded periodic response even without momentum relaxation, and the effect of impurity-induced coupling between the momentum mode and the relaxing modes is suppressed by $\Gamma/\omega$.

For $\omega\ll\Gamma$, impurities instead set the stationary momentum response, whose amplitude grows as $1/\Gamma$. An energy-dependent impurity collision operator generally couples this momentum mode to the modes relaxed by electron-electron collisions. Although the coupling is proportional to the disorder strength, it acts on a momentum response inversely proportional to that strength. Their product therefore remains finite as the disorder is weakened. As shown below, the resulting correction depends on the energy dependence of impurity scattering, but not on its overall magnitude, and changes the dc hydrodynamic anomalous Hall conductivity relative to its clean limit value.

\subsection{Exact orthogonality to the collisional invariants}
\label{sec:orthogonality}

The generation vector \eqref{eq:G} inherits three exact identities from the conservation laws of
the pair collision. Consider the moment $\sum_{\pp}\Lambda(\pp)\,G_{\pp,b}$ for
$\Lambda\in\{1,\varepsilon_{\pp},p_a\}$. Writing the sum over all four legs and symmetrizing with
the relabelings $(1\leftrightarrow2)$, $(3\leftrightarrow4)$, and $(12)\leftrightarrow(34)$,
under the last of which $W$ and, by energy conservation, the equilibrium occupation product are
invariant while $\dr\to-\dr$, one finds
\begin{equation}
\begin{aligned}
\sum_{\pp}\Lambda(\pp)\,G_{\pp,b}
= \frac{1}{8T}\sum_{1234} &W^{34}_{12}\, f_1f_2(1-f_3)(1-f_4)\,\\
&\times\big(\Lambda_1+\Lambda_2-\Lambda_3-\Lambda_4\big)\,\delta r_b .
\end{aligned}
\label{eq:moments}
\end{equation}
For $\Lambda = 1$ the bracket vanishes trivially; for $\Lambda = \varepsilon$ and
$\Lambda = p_a$ it vanishes by energy and momentum conservation. Hence
\begin{equation}
\sum_{\pp}\bm G_{\pp} = 0,\qquad
\sum_{\pp}\varepsilon_{\pp}\,\bm G_{\pp} = 0,\qquad
\sum_{\pp} p_a\,G_{\pp,b} = 0,
\label{eq:conslaws}
\end{equation}
for all $a,b$. The same statement in operator language: the anomalous drive $e\EE\cdot\bm G$ is exactly
orthogonal to the null space $\{1,\pp,\varepsilon\}$ of $\hat C$, so that, in the clean $\omega\to0$ limit,
$\varphi^{g}=-\hat R_{\rm ee}[e\EE\cdot\bm g]$ exists without a momentum-relaxing regulator.
Here $\hat R_{\rm ee}$ denotes the electron-electron relaxation inverse: it inverts
$\hat C$ on the subspace orthogonal to its collision invariants and vanishes on those
invariants. Its explicit first-harmonic construction is given in Sec.~\ref{sec:projection}.
This statement constrains the source; it does not forbid disorder from inducing a momentum
component in the solution of the combined kinetic equation. The momentum zero mode itself follows directly from the fact that the drifted Fermi-Dirac family
$f = \{\exp[(\varepsilon_{\pp}-\bm u\cdot\pp-\mu)/T]+1\}^{-1}$ annihilates every term of the
collision integral (the combination $\ln[f/(1-f)]$ is then a collisional invariant, and the
in/out occupation products cancel identically under the conservation deltas), so the momentum mode
is an exact zero mode of the electron-electron collision operator, and no source built from momentum-
conserving collisions can drive it.

\subsection{Vector structure, energy profile, and the velocity residue}
\label{sec:vector}

By isotropy, $\bm G_{\pp} = G_{\parallel}(\xi)\,\hat\pp + G_{\perp}(\xi)\,\zhat\times\hat\pp$, with
$\xi = \varepsilon_{\pp}-\eF$. The longitudinal component vanishes by an antiunitary mirror
symmetry of the model. Under reflection of all momenta about the $\hat\pp$ axis, the Hamiltonian
\eqref{eq:hamiltonian} is mapped to itself by complex conjugation ($\sigma_y\to-\sigma_y$, $m$ fixed), so
all overlap moduli, and hence $W$, are invariant, while the shift transforms as
$\dr\to -M\dr$, with $M$ the reflection matrix: this follows at once from the explicit form
\eqref{eq:P}, in which $(\bm a-\bm b)\times\zhat \to -M[(\bm a-\bm b)\times\zhat]$ while all scalars
are invariant. Applying the reflection to the collision sum defining $\bm G_{\pp}$ at $\pp$ on the
reflection axis maps the configuration manifold onto itself and gives $\bm G = -M\bm G$, i.e.
$G_{\parallel} = 0$:
\begin{equation}
\begin{aligned}
&\bm G_{\pp} = G_{\perp}(\xi)\;\zhat\times\hat\pp,
\qquad
G_{\perp}(\xi) \;=\; g_\perp(\xi)\,\Big({-}\frac{\partial f_0}{\partial\varepsilon}\Big),\\
&g_\perp(-\xi) = -g_\perp(\xi)
+O\!\left[(T/\eF)\,g_{\rm typ}\right].
\end{aligned}
\label{eq:Gstructure}
\end{equation}
Here $g_{\rm typ}$ denotes a typical magnitude of $g_\perp$ within the thermal shell
$|\xi|\lesssim T$.
The oddness in $\xi$ expresses the particle-hole structure of the source: by
Eq.~\eqref{eq:drFS} the active shift is linear in the energy off-shellness of the pair, and to
leading order in $T/\eF$ the collision kernel is particle-hole symmetric, so the first
non-vanishing contribution to $g_\perp$ is odd under $\xi\to-\xi$.

The low-temperature power counting is fixed by phase space and by the oddness in $\xi$, but the
energy profile does not in general factor from the angular integral. It is useful to write
\begin{equation}
\begin{aligned}
g_\perp(\xi)&=(\nu_F\lambda)^2
\frac{T^3}{\eF p_F^2v_F}\,
\mathcal G\!\left(\frac{\xi}{T},\frac{m}{\eF};\frac{T}{\eF}\right),\\
\mathcal G(-y,r;t)&=-\mathcal G(y,r;t)+O(t),
\end{aligned}
\label{eq:gprofile}
\end{equation}
where $y=\xi/T$, $r=m/\eF$, $t=T/\eF$, and $\mathcal G$ is dimensionless, odd in $m$, and may contain collinear logarithms. If the
angular kernel were smooth on the Fermi surface, the standard Fermi-liquid integrals of
Appendix~\ref{app:profile} would give $\mathcal G\propto y(\pi^2+y^2)$. In two dimensions the
thermal energies also cut off the collinear endpoints, so subleading terms can have additional
$y=\xi/T$ dependence; no polynomial profile is assumed in the numerical collision moments below.

For comparison, the quasiparticle decay rate has the scaling form
\begin{equation}
\frac{1}{\tau_{\rm qp}(\xi)}=(\nu_F\lambda)^2
\frac{T^2}{p_Fv_F}\,
\mathcal R\!\left(\frac{\xi}{T},\frac{m}{\eF};\frac{T}{\eF}\right).
\label{eq:qprate}
\end{equation}
The same thermal phase space occurs in Eqs.~\eqref{eq:gprofile} and \eqref{eq:qprate}.  The quasiparticle decay rate does not, however, determine relaxation of the current-carrying distribution. That relaxation depends on the full collision operator, including its gain terms, and is treated below using the quadratic form in Eq.~\eqref{eq:Q-def}.

Two exact statements now combine to annihilate the naive estimate of the Hall response. First, by
Eq.~\eqref{eq:conslaws} the moments $\sum_{\pp}p_aG_{\pp,b}$ vanish identically. Second, on a single
isotropic Fermi surface the velocity is parallel to the momentum with a common coefficient.
Writing the momentum magnitude and group speed at energy $\varepsilon$ as
$k(\varepsilon)=\sqrt{\varepsilon^2-m^2}/v$ and
$u(\varepsilon)=v^2k(\varepsilon)/\varepsilon$, respectively, with $u(\eF)=v_F$,
we denote the corresponding radial momentum profile by $K(\varepsilon)\equiv k(\varepsilon)$.
Thus $k$ describes the kinematics, while $K$ labels the radial function in the projection
formulas below. The momentum conservation identity~\eqref{eq:conslaws} gives
\begin{equation}
\sum_{\pp} v_{\pp,x}G_{\pp,y}
=\sum_{\pp}\left[u(\varepsilon_{\pp})-\frac{v_F}{p_F}K(\varepsilon_{\pp})\right]
\hat p_xG_{\pp,y},
\end{equation}
where the bracket is $O(v_FT/\eF)$ in the thermal shell. Thus the Fermi-surface part of the
current--generation overlap vanishes; the first surviving term contains the off-shell velocity
residue. In the clean limit, the collision resolvent does not mix momentum with the
relaxing modes, so its mixed Hall matrix element uses only this velocity residue.
Define the radial pairing
\begin{equation}
\langle a|b\rangle_{\rm rad}\equiv\int_{|m|}^{\infty}\!d\varepsilon\,
\nu(\varepsilon)(-\partial_\varepsilon f_0)\,a(\varepsilon)b(\varepsilon),
\end{equation}
where $\nu(\varepsilon)=\varepsilon/(2\pi v^2)$. Functions written in terms of
$\xi$ are evaluated at $\xi=\varepsilon-\eF$. For the cosine harmonic,
$\langle a\cos\vartheta|b\cos\vartheta\rangle=\tfrac12\langle a|b\rangle_{\rm rad}$.
With $\mathsf Q$ denoting the projector introduced in Sec.~\ref{sec:projection},
define the momentum-orthogonal radial velocity by
\begin{equation}
\begin{aligned}
u_{\mathsf Q}(\xi)&\equiv u(\varepsilon)-c_0K(\varepsilon),\\
c_0&\equiv\frac{\langle K|u\rangle_{\rm rad}}{\langle K|K\rangle_{\rm rad}}
=\frac{v_F}{p_F}\left[1+O\!\left(\frac{T^2}{\eF^2-m^2}\right)\right].
\end{aligned}
\end{equation}
Using $du/d\varepsilon = m^2/(\varepsilon^2 K)$ and
$dK/d\varepsilon = \varepsilon/(v^2K)$, one may write
\begin{equation}
\begin{aligned}
u_{\mathsf Q}(\xi) &= \xi\Big[\frac{du}{d\varepsilon}-\frac{v_F}{p_F}\frac{dK}{d\varepsilon}\Big]
+O\!\left[v_F\frac{\xi^2+T^2}{\eF^2-m^2}\right]\\
&= -\frac{\xi}{p_F}\Big(1-\frac{m^2}{\eF^2}\Big)
+O\!\left[v_F\frac{\xi^2+T^2}{\eF^2-m^2}\right]\\
&= -\frac{\xi}{p_F}\Big(1-\frac{p_F}{v_F}\frac{\partial v_F}{\partial p_F}\Big)
+O\!\left[v_F\frac{\xi^2+T^2}{\eF^2-m^2}\right].
\end{aligned}
\label{eq:vperp}
\end{equation}
The bracket is the particle-hole (nonparabolicity) asymmetry factor of the band: it
vanishes identically for a parabolic dispersion ($u\propto p$), for which a rigidly drifting Fermi
sea carries the momentum sector and the projected clean interparticle charge Hall response
vanishes.
For the Dirac band the asymmetry factor equals $v^2p_F^2/\eF^2$. The clean Hall response is therefore suppressed by two powers of $T/\eF$, one from
$u_{\mathsf Q}\sim\xi\sim T$ in the current vertex and one from the $\xi$-oddness of $g_\perp$
in the drive. After inversion of the collision operator this gives the explicit $T^2$ factor
derived in Sec.~\ref{sec:hplot}, multiplied by the dimensionless endpoint-dependent coefficient $\hc$.
With disorder, the velocity entering the response must first be obtained from the momentum balance.

\subsection{Momentum relaxation and the effective driving velocity}
\label{sec:projection}

Work in the cosine harmonic and denote the momentum mode by
$\Phi_0(\pp)\equiv p_x=K(\varepsilon_{\pp})\cos\vartheta$. Define the momentum projector and its
complement by
\begin{equation}
\mathsf P=\frac{|\Phi_0\rangle\langle\Phi_0|}{\mathcal N},\qquad
\mathsf Q=1-\mathsf P,\qquad \mathcal N=\langle\Phi_0|\Phi_0\rangle,
\label{eq:momentum-projectors}
\end{equation}
and introduce
\begin{equation}
\begin{aligned}
\mathcal V&=\langle\Phi_0|v_x\rangle,\qquad
\mathcal D=\langle\Phi_0|\hat D|\Phi_0\rangle,\\
\Gamma&=\mathcal D/\mathcal N,\\
|d\rangle&=\mathsf Q\hat D|\Phi_0\rangle,\qquad
\hat C_{\mathsf Q}=\hat C\big|_{\mathsf Q}.
\end{aligned}
\label{eq:momentum-moments}
\end{equation}
The moments $\mathcal N$, $\mathcal V$, and $\mathcal D$ are scalars obtained by integrating
over momentum; they depend on the model parameters and temperature, but not on a free $\pp$.
The radial projection above obeys $c_0=\mathcal V/\mathcal N$ and
$\mathsf Q|v_x\rangle=|u_{\mathsf Q}\cos\vartheta\rangle$.
The generation vector is in $\mathsf Q$ by Eq.~\eqref{eq:conslaws}.
To solve the ordinary drive, set $\varphi^v=a_v\Phi_0+\psi_v$, where
$\langle\Phi_0|\psi_v\rangle=0$. The two projections of the Boltzmann equation for $E_x$ are
\begin{equation}
\begin{aligned}
(\mathcal D-i\omega \mathcal N)a_v+\langle d|\psi_v\rangle&=eE_x\mathcal V,\\
(\hat C_{\mathsf Q}+\mathsf Q\hat D\mathsf Q-i\omega)\psi_v+a_vd&=eE_x\mathsf Qv_x.
\end{aligned}
\label{eq:projected-BE}
\end{equation}
Eliminating $a_v$ gives an effective drive
\begin{equation}
|v_{{\rm eff},\omega}\rangle
=\mathsf Q|v_x\rangle
-\frac{\mathcal V|d\rangle}{\mathcal N(\Gamma-i\omega)}.
\label{eq:effective-velocity}
\end{equation}
The exact reduced operator is given in Appendix~\ref{app:disorder-projection}.
For $\{\omega,\Gamma\}\ll\gamma_{\rm ee}$ it can be replaced by $\hat C_{\mathsf Q}$ in the relaxing sector,
with perturbative corrections controlled by $(\Gamma+\omega)/\gamma_{\rm ee}$.
Equation~\eqref{eq:sigma-2x} then becomes
\begin{equation}
\sigma_{xy}^{\rm shift}(\omega)
\simeq-2e^2\langle v_{{\rm eff},\omega}|\hat R_{\rm ee}|g_y\rangle.
\label{eq:hydro-bilinear}
\end{equation}
In this harmonic the relaxation inverse is explicitly
\begin{equation}
\hat R_{\rm ee}\equiv\mathsf Q\hat C_{\mathsf Q}^{-1}\mathsf Q,\qquad
\hat C\hat R_{\rm ee}=\hat R_{\rm ee}\hat C=\mathsf Q,
\end{equation}
with $\hat R_{\rm ee}|\Phi_0\rangle=0$. Thus only the relaxing subspace is inverted.
Unlike the full resolvent $\hat R_\omega$, $\hat R_{\rm ee}$ contains neither disorder nor
frequency. The pairing remains bilinear when $v_{{\rm eff},\omega}$ is complex.

The order of these operations matters. In dc, $a_v$ scales as $\Gamma^{-1}$, while $d$ scales
as $\Gamma$. Thus $a_vd$ survives as the disorder strength tends to zero. One may neglect
$\hat D$ as an additional relaxation rate on $\mathsf Q$ only after eliminating momentum;
otherwise the second term in Eq.~\eqref{eq:effective-velocity} is lost.

The anomalous distribution also acquires a momentum component. With
$\varphi^g=a_g\Phi_0+\psi_g$ and field $E_y$, its projected momentum equation is
\begin{equation}
a_g=-\frac{\langle d|\psi_g\rangle}{\mathcal N(\Gamma-i\omega)}.
\label{eq:anomalous-drift}
\end{equation}
Its ordinary current therefore obeys
\begin{equation}
j_x^{\rm ad}=e\big[\langle v_x|\psi_g\rangle+\mathcal V a_g\big]
=e\langle v_{{\rm eff},\omega}|\psi_g\rangle.
\label{eq:anomalous-effective-current}
\end{equation}
Even though its source injects no momentum, $a_g$ has a finite disorder-selected dc limit.
This contribution is why disorder changes both halves of the Hall current equally, rather
than changing shift accumulation alone.

\section{The hydrodynamic Hall conductivity}
\label{sec:conductivity}

\subsection{Self-regularization of the contact model}
\label{sec:regularization}

The calculation is organized for the contact interaction $U = \lambda$, for three reasons. First,
the exact factorization \eqref{eq:factor} gives the shift and the scattering probability in closed
form, Eqs.~\eqref{eq:P}, \eqref{eq:R}, and \eqref{eq:W2}. Second, the collinear structure of the
shift is under control within the model: the pair function is bounded [item (c) of
Appendix~\ref{sec:closed}], and every matrix element of transport theory involves the product
$W\,\dr$ of Eq.~\eqref{eq:Wdr}, in which the Pauli factor
$(1-\bm n_{\kk_1}\!\cdot\bm n_{\kk_2})$ [respectively $(1-\bm n_{\pp_1}\!\cdot\bm n_{\pp_2})$]
multiplies the denominator of the initial-pair (final-pair) term and suppresses precisely the
configurations in which the pair function is largest, so no external screening cutoff is needed:
the spinor structure of the chiral band regularizes the theory by itself. Third, weak coupling $\nu_F\lambda\ll1$ controls the Born collision kernel. A hydrodynamic
transport regime additionally requires momentum relaxation to be slower than the relevant
electron--electron relaxation rates. As shown below, the common factor $\lambda^2$ drops out of the coordinate-shift answer at leading
Born order.

\subsection{Galerkin scheme}
\label{sec:scheme}

We use $\hc_{\rm cl}\equiv\hc$ and $\hc_{\rm dc}$ for the clean limit and dc
coefficients, respectively, with $\sigma^a_{xy}=(e^2/\hbar)t^2\hc_a$ and $a\in\{\mathrm{cl},\mathrm{dc}\}$.
Their finite-$T$ values retain the full thermal width. We denote a clean leading-Sommerfeld
(LS) estimate in an $n$-mode odd radial basis by $\hc_{\rm LS}^{(n)}$; its collision
moments retain the exact off-shell kinematics, while its drive is expanded at leading
Sommerfeld order. The cases $n=1$ and $n=3$ are compared below.

The leading hydrodynamic conductivity \eqref{eq:hydro-bilinear} requires the inverse of $\hat C$
on the first angular harmonic, restricted to the complement of the momentum zero mode
$\Phi_0(\pp)=K(\varepsilon_{\pp})\cos\vartheta$. The same inverse and generation vector apply throughout the
frequency crossover; only the effective velocity depends on $\omega/\Gamma$.
The projection must be made explicitly at finite temperature.  For any raw radial trial function
$\chi(\xi)$ define
\begin{equation}
 \bar\chi(\xi)\equiv \chi(\xi)-c[\chi] K(\varepsilon),
 \qquad
 c[\chi]\equiv\frac{\langle K|\chi\rangle_{\rm rad}}{\langle K|K\rangle_{\rm rad}},
 \label{eq:chi-project}
\end{equation}
so that $\langle K|\bar\chi\rangle_{\rm rad}=0$. The effective velocity of Eq.~\eqref{eq:effective-velocity} and, by
Eq.~\eqref{eq:conslaws}, the generation term live in this complement.  We approximate the inverse
there by the one-mode expression
\begin{equation}
\hat R_{\rm ee}\simeq
\frac{|\bar\chi\cos\vartheta\rangle\langle\bar\chi\cos\vartheta|}
{\langle\bar\chi\cos\vartheta|\hat C|\bar\chi\cos\vartheta\rangle},
\end{equation}
and use, at leading Sommerfeld order, the signed linear velocity residue
\begin{equation}
\chi_1(\xi)=-\frac{\xi}{p_F}\Big(1-\frac{m^2}{\eF^2}\Big)
=u_{\mathsf Q}(\xi)+O\!\left[v_F\frac{\xi^2+T^2}{\eF^2-m^2}\right].
\label{eq:trial}
\end{equation}
In the collision differences entering $J[\chi]$ and $Q[\chi]$, the subtraction in
Eq.~\eqref{eq:chi-project} cancels exactly because
$\Delta[K(\varepsilon)\hat\pp]=\Delta\pp=0$.  It does not cancel from the drive overlap;
there the projected function $\bar\chi$ must be used.  The one-mode inversion gives~\footnote{For
a Hall response, with $\hat R_{\rm ee}$
sandwiched between two different vectors, the one-mode truncation is a finite-dimensional approximation
rather than a variational bound, and the higher odd modes $\xi^3,\xi^5,\ldots$ contribute at the
same thermal order, so its accuracy must be checked rather than assumed. The trial shape
$\chi\propto\xi$ is the lowest smooth odd mode and the leading velocity residue; enlarging the
trial space to $\{\xi,\xi^3,\xi^5\}$ changes the coefficient by less than one percent
(Sec.~\ref{sec:hplot}), and the numerical integrals were checked with independent phase-space parametrizations.}
\begin{equation}
\sigma^{\rm shift}_{xy}(\omega) \;\simeq\; -\,\frac{2e^2\,D[\chi;\omega]\,J[\chi]}{Q[\chi]}.
\label{eq:sigma-var}
\end{equation}
Here $D[\chi;\omega]=\langle\bar\chi\cos\vartheta|v_{{\rm eff},\omega}\rangle$.
The brackets emphasize that these moments are scalar functionals of the trial function;
they have no free momentum argument. We reserve $D[\chi]$ for the clean-limit value
and define the three common moments
(the clean drive overlap, generation overlap, and entropy-production form) by
\begin{align}
D[\chi] &\equiv \big\langle\bar\chi\cos\vartheta\big|u_{\mathsf Q}\cos\vartheta\big\rangle
\nonumber\\
&= \tfrac12\!\int\!d\xi\;\nu(\varepsilon)\,\bar\chi(\xi)\,u_{\mathsf Q}(\xi)
\Big({-}\frac{\partial f_0}{\partial\xi}\Big),
\label{eq:D-def}\\[2pt]
J[\chi] &\equiv \big\langle\chi\cos\vartheta\big|g_\perp\cos\vartheta\big\rangle
\nonumber\\
&= \sum_{\pp}\chi(\xi_{\pp})\hat p_xG_{\pp,y}
\nonumber\\
&= \frac{1}{8T}\sum_{1234}W^{34}_{12}\,\mathscr F_{1234}\,
\big[\Delta(\chi\hat\pp)\big]_x\,\delta r_y
\nonumber\\
&= \frac{1}{16T}\sum_{1234}W^{34}_{12}\,\mathscr F_{1234}\,
\big[\Delta(\chi\hat\pp)\times\dr\big]_z ,
\label{eq:J-def}\\[2pt]
Q[\chi] &\equiv \big\langle\chi\cos\vartheta\big|\hat C
\big|\chi\cos\vartheta\big\rangle
\nonumber\\
&= \frac{1}{8T}\sum_{1234}W^{34}_{12}\,\mathscr F_{1234}\,
\big[\Delta(\chi\hat\pp)\big]_x^2
\nonumber\\
&= \frac{1}{16T}\sum_{1234}W^{34}_{12}\,\mathscr F_{1234}\,
\big|\Delta(\chi\hat\pp)\big|^2 ,
\label{eq:Q-def}
\end{align}
with $\nu(\varepsilon)=\varepsilon/(2\pi v^2)$ the conduction-band density of states,
$\mathscr F_{1234}\equiv f_1f_2(1{-}f_3)(1{-}f_4)$, and
$\Delta(\chi\hat\pp) \equiv \chi(\xi_1)\hat\pp_1+\chi(\xi_2)\hat\pp_2
-\chi(\xi_3)\hat\pp_3-\chi(\xi_4)\hat\pp_4$. The last lines of Eqs.~\eqref{eq:J-def} and
\eqref{eq:Q-def} follow by averaging over a global rotation of all four momenta,
$\langle\Delta_x\delta r_y\rangle = \tfrac12(\bm\Delta\times\dr)_z$ and
$\langle\Delta_x^2\rangle = \tfrac12|\bm\Delta|^2$; they are the forms to be used whenever the
integration exploits isotropy by fixing the direction of one momentum, and the same reduction
must be applied to both moments~\footnote{Mixing the unaveraged Cartesian expression
for $J[\chi]$ with the rotationally averaged expression for $Q[\chi]$ in a frame with one
momentum fixed overestimates $J[\chi]/Q[\chi]$ by roughly $40\%$ in the present
parametrization. This is an angular-reduction artifact.}.

For a finite raw radial basis $\{\chi_a\}$, project every member with
Eq.~\eqref{eq:chi-project} and define
$D_a(\omega)=D[\chi_a;\omega]$, $J_a=J[\chi_a]$, and
$Q_{ab}=\langle\chi_a\cos\vartheta|\hat C|\chi_b\cos\vartheta\rangle$.
Here $D[\chi;\omega]$ and its basis components $D_a(\omega)$ denote drive overlaps;
$\hat D$ is the impurity collision operator and $\mathcal D$ its momentum-mode matrix element.
Conservation makes
$J_a$ and $Q_{ab}$ unchanged by the projection, and the consistent Galerkin result is
\begin{equation}
 \sigma_{xy}^{\rm shift}(\omega)\simeq-2e^2\bm D(\omega)^{\mathsf T}Q^{-1}\bm J.
 \label{eq:sigma-shift-multimode}
\end{equation}
In the clean limit, $D_a^{\rm cl}=D[\chi_a]$. This specialization
is used for the three-mode check below; the dc comparison changes the drive vector without
recomputing $Q$ or $\bm J$.  At strictly leading
particle--hole-symmetric order the odd basis $\{\xi,\xi^3,\xi^5\}$ is closed; a full finite-$T$
Galerkin calculation must also allow even radial modes.

For the exact choice $\chi=u_{\mathsf Q}$, one has $\bar\chi=\chi$ and
$D[\chi]=\langle u_{\mathsf Q}\cos\vartheta|u_{\mathsf Q}\cos\vartheta\rangle>0$. Using the radial
pairing defined in Sec.~\ref{sec:vector},
\begin{equation}
 D[u_{\mathsf Q}]=\frac12\left[\langle u|u\rangle_{\rm rad}
 -\frac{\langle K|u\rangle_{\rm rad}^{\,2}}{\langle K|K\rangle_{\rm rad}}\right]>0.
 \label{eq:D-exact}
\end{equation}
For the leading trial
\eqref{eq:trial}, Eq.~\eqref{eq:D-def} is evaluated after the projection
\eqref{eq:chi-project}; its leading Sommerfeld term is the same positive norm, while regular
finite-$T$ corrections are kept separate.  Moreover,
$\sum_{\pp}p_xG_{\pp,y}=0$, so replacing $u$ by $u_{\mathsf Q}$ in the direct current--generation overlap is
exact. Inside the disorder-dressed response the required vertex is instead Eq.~\eqref{eq:effective-velocity}. Under an arbitrary
rescaling $\chi\to a\chi$, $D[a\chi]=aD[\chi]$, $J[a\chi]=aJ[\chi]$, and
$Q[a\chi]=a^2Q[\chi]$; hence $D[\chi] J[\chi]/Q[\chi]$ is invariant, including under $a=-1$.
Retaining the sign of $u_{\mathsf Q}$ in Eq.~\eqref{eq:trial} makes $D[\chi]>0$ and places the physical
sign in $J[\chi]$. Both $J[\chi]$ and $Q[\chi]$ are
conditioned by the on-shell cancellations of Appendix~\ref{sec:dirac} and Sec.~\ref{sec:conservation}:
their integrands vanish pointwise on the Fermi surface [$\dr = 0$ and $\Delta(\chi\hat\pp) = 0$
there] and are quadratic in the thermal off-shellness. The linear trial function is the lowest
smooth odd radial mode and coincides with the leading velocity residue; its accuracy for the
collision inverse is assessed by the multimode stability test rather than by this power
counting alone.

\subsection{The clean limit}
\label{sec:hplot}

In the clean limit, $\Gamma\ll \omega \ll\gamma_{\rm ee}$, the momentum-feedback term in Eq.~\eqref{eq:effective-velocity} is suppressed
by $\Gamma/\omega$, so $v_{{\rm eff},\omega}\to u_{\mathsf Q}\cos\vartheta$. The limiting coefficient is
\begin{equation}
\sigma_{xy}^{\rm cl}=-2e^2\langle u_{\mathsf Q}\cos\vartheta|\hat R_{\rm ee}|g_y\rangle.
\label{eq:clean-limit}
\end{equation}
The superscript ``cl'' specifies the clean limit; the unadorned coefficient $\hc$ below
refers to this clean endpoint. The limit takes momentum relaxation to zero before frequency.
The longitudinal distribution has a momentum pole, but the clean mixed Hall matrix element
has none because $\langle\Phi_0|g_y\rangle=0$. Thus the regular clean Hall response remains
finite even though a clean homogeneous metal has no stationary longitudinal drift.

The low-temperature normalization of the three moments \eqref{eq:D-def}--\eqref{eq:Q-def} is
specified in Appendix~\ref{app:sommerfeld}, Eqs.~\eqref{eq:Dresult}--\eqref{eq:Qresult}. Using the clean value $D[\chi]$ in
Eq.~\eqref{eq:sigma-var}, three cancellations occur, each with physical content:
\emph{(i)} the coupling $(\nu_F\lambda)^2$ cancels between $J[\chi]$ and $Q[\chi]$: generation and
relaxation are provided by the same collisions; \emph{(ii)} both exact collision moments scale as
$T^4$ times dimensionless endpoint functions, so their common thermal power cancels and leaves the
$T^2$ of $D[\chi]$; \emph{(iii)} one power of the asymmetry factor
$(1-m^2/\eF^2)$ cancels in the leading-Sommerfeld (LS) one-mode estimator.
The resulting conductivity is
\begin{equation}
\begin{aligned}
\sigma^{\rm cl}_{xy}\simeq{}&\frac{\pi^2e^2}{3}\,\nu_FT^2
\frac{1-m^2/\eF^2}{p_F^2\eF}\frac{\mathcal A(m/\eF;T/\eF)}
{\mathcal B(m/\eF;T/\eF)}
\end{aligned}.
\label{eq:sigmaAB}
\end{equation}
Using $\nu_F=\eF/(2\pi v^2)$ and $v^2p_F^2=\eF^2-m^2$, the remaining kinematic
prefactors satisfy $\nu_F(1-m^2/\eF^2)/(p_F^2\eF)=1/(2\pi\eF^2)$.
Thus, writing $r=m/\eF$ and $t=T/\eF$ and restoring $\hbar$, Eq.~\eqref{eq:sigmaAB} reduces to
\begin{equation}\label{eq:hmain}
\sigma^{\rm cl}_{xy}\simeq\hc(r;t)\frac{e^2}{\hbar}t^2,\quad 
\hc_{\rm LS}^{(1)}(r;t)=\frac{\pi}{6}\frac{\mathcal A(r;t)}{\mathcal B(r;t)}.
\end{equation}
The first relation defines the clean coefficient for any consistent Galerkin inversion. In the
LS one-mode inversion, it reduces to $\pi/6$ times the ratio of two dimensionless
moments: the chirality-weighted generation amplitude $\mathcal A$, defined by the exact
$J[\chi]$, and the mode-relaxation amplitude $\mathcal B$, defined by the exact $Q[\chi]$. We
suppress the residual temperature argument of $\hc$ when quoting results at a fixed temperature.
As specified in Sec.~\ref{sec:scheme}, the LS collision moments retain the exact
off-shell kinematics at $t>0$ to resolve the collinear cutoff. The one-mode estimate
$\hc_{\rm LS}^{(1)}$ is distinct from the converged finite-$t$ Galerkin calculation
in the mixed-parity basis.
The moment amplitudes obey $\mathcal A(-r;t)=-\mathcal A(r;t)$ and
$\mathcal B(-r;t)=\mathcal B(r;t)>0$, so $\hc_{\rm LS}^{(1)}$ is odd; the full projected result has
the same parity
in $m$. The computed mass dependence is approximately linear over the range specified below;
oddness alone does not determine its slope. To leading Born order, the projected clean result
is independent of interaction strength. As discussed in
Sec.~\ref{sec:logs}, $\hc_{\rm LS}^{(1)}$ varies by less than eight percent at the two representative masses over the extended computed temperature
window; the asymptotic collinear dependence of the ratio remains unresolved. The finite-temperature
collision integral gives $\mathcal A>0$ for $m>0$, hence $\hc_{\rm LS}^{(1)}>0$; the converged
multimode Galerkin values are also positive. Thus the sign of the collisional term
is opposite to that of the intrinsic contribution of the same cone,
$\sigma^{\rm int}_{xy} = -(e^2/4\pi\hbar)(m/\eF)$ in the convention used throughout this paper
[$j_i = \sigma_{ij}E_j$; anomalous velocity
$\dot\rr_{\rm an}=-\dot\kk\times[\Omega(\kk)\zhat]$
with $\hbar\dot\kk=e\EE$, which fixes the intrinsic sign used here]: the computed collisional term opposes the intrinsic contribution. The temperature
dependence of their sum also includes the intrinsic Sommerfeld term (Sec.~\ref{sec:summary}).

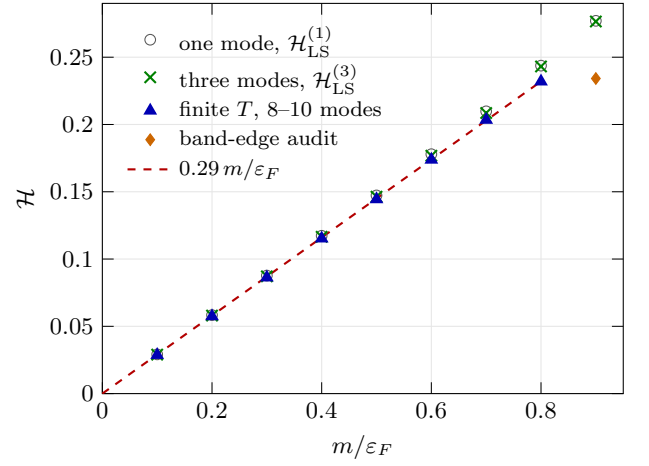
\begin{figure}[!htbp]
\centering
\begin{tikzpicture}
\begin{axis}[width=0.98\columnwidth,height=0.78\columnwidth,
xlabel={$m/\eF$},
ylabel={$\hc$},
xmin=0,xmax=0.95,ymin=0,ymax=0.29,
legend pos=north west,legend style={draw=none,fill=none,font=\footnotesize},
legend cell align=left,scaled y ticks=false,
yticklabel style={/pgf/number format/fixed,/pgf/number format/precision=2},
tick style={black},grid=both,grid style={gray!20}]
\addplot[only marks,mark=o,mark size=2.0pt,color=gray!70!black] coordinates {
(0.1,0.0291493) (0.2,0.0583613) (0.3,0.0877096) (0.4,0.1172941) (0.5,0.1472750)
(0.6,0.1779193) (0.7,0.2097806) (0.8,0.2437267) (0.9,0.2770405)};
\addlegendentry{one mode, $\hc_{\rm LS}^{(1)}$}
\addplot[only marks,mark=x,mark size=3.0pt,thick,color=green!50!black] coordinates {
(0.1,0.0289623) (0.2,0.0579862) (0.3,0.0871443) (0.4,0.1165356) (0.5,0.1463237)
(0.6,0.1767809) (0.7,0.2085310) (0.8,0.2431192) (0.9,0.2765362)};
\addlegendentry{three modes, $\hc_{\rm LS}^{(3)}$}
\addplot[only marks,mark=triangle*,mark size=2.2pt,thick,color=blue!70!black] coordinates {
(0.1,0.0287194) (0.2,0.0574769) (0.3,0.0863174) (0.4,0.1152830)
(0.5,0.1444389) (0.6,0.1738454) (0.7,0.2034681) (0.8,0.2319524)};
\addlegendentry{finite $T$, 8--10 modes}
\addplot[only marks,mark=diamond*,mark size=2.2pt,color=orange!80!black] coordinates {(0.9,0.2341726)};
\addlegendentry{band-edge audit}
\addplot[dashed,thick,domain=0:0.8,color=red!70!black] {0.29*x};
\addlegendentry{$0.29\,m/\eF$}
\end{axis}
\end{tikzpicture}
\caption{Mass dependence of the clean limit hydrodynamic shift coefficient of
Eq.~\eqref{eq:hmain}, $\sigma^{\rm cl}_{xy} \simeq
\hc(m/\eF;T/\eF)(e^2/\hbar)(T/\eF)^2$ per cone, for the contact interaction at $T/\eF=0.02$. Open circles: one-mode
evaluation using the exact off-shell collision moments $J[\chi],Q[\chi]$. Crosses: the corresponding
three-component generation vector and $3\times3$ collision matrix for the enlarged odd radial
space $\{\xi,\xi^3,\xi^5\}$. Both use the exactly momentum-projected basis
\eqref{eq:chi-project} and the leading Sommerfeld drive vector.
Their separation therefore tests
basis convergence at leading low-temperature order, not the full finite-$T$ systematic error.
Triangles: the finite-$T$ Galerkin result with exact drive projection and a converged mixed-parity
basis (eight modes through $m/\eF=0.7$ and ten modes at $0.8$); the orange diamond is the ten-mode
$m/\eF=0.9$ band-edge audit. Dashed line:
the approximate linear dependence \eqref{eq:hlinear}, shown only through $m/\eF=0.8$. The coefficient vanishes with
the Berry curvature at $m\to0$. The separation of the LS and finite-$T$ series near the band edge
is the expected breakdown of the fixed-$r$ Sommerfeld expansion, not a sampling uncertainty.}
\label{fig:h}
\end{figure}

The amplitudes $\mathcal A$ and $\mathcal B$ are normalized exact collision moments; their ratio
is evaluated by computing the full collision integrals
\eqref{eq:J-def}--\eqref{eq:Q-def}, in their rotationally reduced forms, with the exact
off-shell kinematics and the regularized product $|V|^2\dr$ obtained from
Eqs.~\eqref{eq:W2} and \eqref{eq:P}. The result is shown in
Fig.~\ref{fig:h}. Over the range $0.1\leq m/\eF\leq0.8$ the converged finite-$T$ coefficient is remarkably
linear in the gap parameter,
\begin{equation}
\hc\Big(\frac{m}{\eF};0.02\Big) \;\simeq\; 0.29\;\frac{m}{\eF},
\qquad 0.1\leq\frac{|m|}{\eF}\leq0.8,
\label{eq:hlinear}
\end{equation}
with only weak curvature.
The linearity reflects the $m$-oddness of the
source (one factor $m/\eF$ from the chirality of $\dr$), with the remaining $m$-dependence of
the projected collision moments compensating almost completely; $\hc\to0$ at $m\to0$, as required, since
the shift vanishes identically with the Berry curvature. Within the leading low-temperature
(Sommerfeld) projection, enlarging the odd radial trial space from $\{\xi\}$ to
$\{\xi,\xi^3,\xi^5\}$, giving $\hc_{\rm LS}^{(3)}$, lowers the coefficient by $0.6\%$--$0.7\%$ for
$0.1\le m/\eF\le0.7$. This tests convergence of the collision inverse within the leading
particle-hole-odd sector; it does not test regular finite-$T$ corrections to the projected
velocity or subleading radial structures. We therefore also performed an enlarged, converged finite-$T$
Galerkin calculation: the exact one-particle metric and projection \eqref{eq:chi-project} are used,
and the radial space contains both parities. The resulting converged values (eight modes through
$m/\eF=0.7$ and ten modes at $0.8,0.9$) are shown in Fig.~\ref{fig:h}. The temperature dependence of the LS collision-moment ratio is examined over a wider range in
Sec.~\ref{sec:logs}.

\subsection{Collinear endpoints and the stability of the Hall ratio}
\label{sec:logs}

Resolving the two angular energy-conservation constraints at fixed momentum transfer $q$
produces the familiar two-dimensional endpoint Jacobians. Their singularity cannot, however, be
counted independently of the contact-interaction matrix element and of the collision vertices.
On the Fermi surface Eq.~\eqref{eq:W2} gives
\begin{equation}
 |V|^2\propto
 \left(1-\frac{m^2}{\eF^2}\right)^2
 (1-\hat\kk_1\!\cdot\hat\kk_2)
 (1-\hat\pp_1\!\cdot\hat\pp_2).
\label{eq:collinear-form-factor}
\end{equation}
In the intermediate regime $|\varpi|/v_F\ll q\ll p_F$, where $\varpi$ is the
collision energy transfer, the near-Fermi-surface angular constraints have co-directed
and counterpropagating roots. This classification is not uniform down to the kinematic
threshold $q\sim|\varpi|/v_F$, where the roots merge. On the co-directed root both factors in Eq.~\eqref{eq:collinear-form-factor} are $O(q^2/p_F^2)$;
therefore $|V|^2=O(q^4/p_F^4)$ and the nominal $dq/q$ singularity is removed. On the
counterpropagating root the form factor remains finite, and thermal energy transfer $\varpi$ cuts off the
remaining endpoint at $q\sim\max(|\varpi|,|\xi_i|)/v_F$, allowing a single logarithm. Near
$q=2p_F$ the matrix element is finite for a head-on pair. Although both
$\Delta(\chi\hat\pp)$ and $\dr$ vanish everywhere on the common shell, their leading off-shell
terms can combine with the endpoint Jacobian; its coefficient must be obtained from their joint
expansion rather than from the Jacobian alone.

At the level of this root-by-root power counting, each endpoint allows at most one logarithm. Adding the two
endpoint contributions changes its coefficient; it does not produce a square of the logarithm.
In the asymptotic domain $T\ll p_Fv_F$, write $L=\ln(p_Fv_F/T)$. A possible leading
low-temperature structure of the exact moments is
\begin{equation}
\begin{aligned}
\mathcal A&=a_1(m/\eF)L+a_0(m/\eF)+\cdots,\\
\mathcal B&=b_1(m/\eF)L+b_0(m/\eF)+\cdots,
\end{aligned}
\label{eq:collinear-expansion}
\end{equation}
where a coefficient may vanish after the full sum over roots. Establishing these coefficients
requires evaluating Eqs.~\eqref{eq:J-def} and \eqref{eq:Q-def} with the same kinematics and angular
reduction.
These endpoint structures belong to the same family as the two-dimensional quasiparticle and
odd-harmonic relaxation anomalies of Refs.~\cite{ledwith19a,ledwith19b}; those results do not,
however, determine the coefficients $a_i,b_i$ for the present vertices.

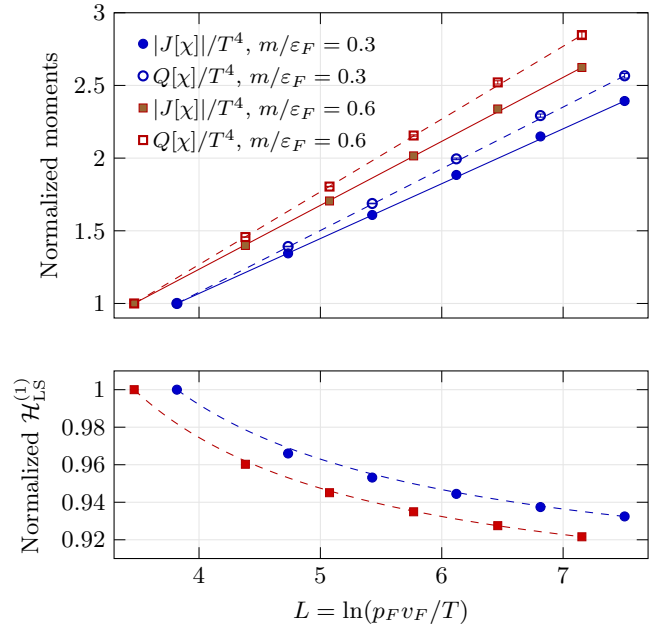
\begin{figure}[!htbp]
\centering
\begin{tikzpicture}
\begin{axis}[name=upper,width=1.0\columnwidth,height=0.65\columnwidth,
 xlabel={},xticklabels={},ylabel={Normalized moments},
 xmin=3.3,xmax=7.7,ymin=0.9,ymax=3.0,
 legend pos=north west,legend style={draw=none,fill=none,font=\footnotesize},
 legend cell align=left,tick style={black},grid=both,grid style={gray!20}]
\addplot+[only marks,mark=*,mark size=1.7pt,color=blue!70!black,error bars/.cd,y dir=both,y explicit] coordinates {(7.506592,2.392829)+-(0,.014960) (6.813445,2.149126)+-(0,.010344) (6.120297,1.884102)+-(0,.005160) (5.427150,1.608805)+-(0,.002198) (4.734003,1.344644)+-(0,.001689) (3.817712,1)+-(0,0)};
\addlegendentry{$|J[\chi]|/T^4$, $m/\eF=0.3$}
\addplot+[only marks,mark=o,mark size=1.7pt,thick,color=blue!70!black,error bars/.cd,y dir=both,y explicit] coordinates {(7.506592,2.566205)+-(0,.015704) (6.813445,2.292557)+-(0,.010936) (6.120297,1.994889)+-(0,.005474) (5.427150,1.687860)+-(0,.001900) (4.734003,1.391976)+-(0,.001689) (3.817712,1)+-(0,0)};
\addlegendentry{$Q[\chi]/T^4$, $m/\eF=0.3$}
\addplot+[only marks,mark=square*,mark size=1.5pt,color=red!70!black,error bars/.cd,y dir=both,y explicit] coordinates {(7.154615,2.623438)+-(0,.020012) (6.461468,2.338121)+-(0,.011131) (5.768321,2.015327)+-(0,.004273) (5.075174,1.705236)+-(0,.000870) (4.382027,1.399161)+-(0,.000569) (3.465736,1)+-(0,0)};
\addlegendentry{$|J[\chi]|/T^4$, $m/\eF=0.6$}
\addplot+[only marks,mark=square,mark size=1.5pt,thick,color=red!70!black,error bars/.cd,y dir=both,y explicit] coordinates {(7.154615,2.846623)+-(0,.021359) (6.461468,2.520820)+-(0,.011259) (5.768321,2.155624)+-(0,.004548) (5.075174,1.804313)+-(0,.001182) (4.382027,1.457084)+-(0,.000708) (3.465736,1)+-(0,0)};
\addlegendentry{$Q[\chi]/T^4$, $m/\eF=0.6$}
\addplot[thin,color=blue!70!black,domain=3.817712:7.506592] {1+0.3775750*(x-3.817712)};
\addplot[thin,dashed,color=blue!70!black,domain=3.817712:7.506592] {1+0.4245746*(x-3.817712)};
\addplot[thin,color=red!70!black,domain=3.465736:7.154615] {1+0.4400899*(x-3.465736)};
\addplot[thin,dashed,color=red!70!black,domain=3.465736:7.154615] {1+0.5005919*(x-3.465736)};
\end{axis}
\begin{axis}[name=lower,at={(upper.south west)},anchor=north west,yshift=-7mm,
 width=1.0\columnwidth,height=0.47\columnwidth,xlabel={$L=\ln(p_Fv_F/T)$},
 ylabel={Normalized $\hc_{\rm LS}^{(1)}$},xmin=3.3,xmax=7.7,ymin=0.91,ymax=1.01,
 tick style={black},grid=both,grid style={gray!20}]
\addplot+[only marks,mark=*,mark size=1.7pt,color=blue!70!black,error bars/.cd,y dir=both,y explicit] coordinates {(7.506592,.932438)+-(0,.001016) (6.813445,.937435)+-(0,.000196) (6.120297,.944464)+-(0,.000478) (5.427150,.953162)+-(0,.000577) (4.734003,.965997)+-(0,.000093) (3.817712,1)+-(0,0)};
\addplot+[only marks,mark=square*,mark size=1.5pt,color=red!70!black,error bars/.cd,y dir=both,y explicit] coordinates {(7.154615,.921594)+-(0,.000207) (6.461468,.927521)+-(0,.000348) (5.768321,.934916)+-(0,.000519) (5.075174,.945089)+-(0,.000198) (4.382027,.960247)+-(0,.000116) (3.465736,1)+-(0,0)};
\addplot[thin,dashed,color=blue!70!black,domain=3.817712:7.506592,samples=80] {(1+0.3775750*(x-3.817712))/(1+0.4245746*(x-3.817712))};
\addplot[thin,dashed,color=red!70!black,domain=3.465736:7.154615,samples=80] {(1+0.4400899*(x-3.465736))/(1+0.5005919*(x-3.465736))};
\end{axis}
\end{tikzpicture}
\caption{Finite-window collinear enhancement of the collision moments and relative stability of
the Hall ratio. Upper panel: $|J[\chi]|/T^4$ (filled symbols) and $Q[\chi]/T^4$ (open symbols),
normalized separately to their values at $T_0=0.02\,\eF$, versus $L=\ln(p_Fv_F/T)$ for
$m/\eF=0.3$ (circles) and $0.6$ (squares). Thin straight lines are chords through
the first and last points, not asymptotic fits. Lower panel:
$\hc_{\rm LS}^{(1)}(T)/\hc_{\rm LS}^{(1)}(T_0)$ from the same one-mode
correlated estimates. Dashed curves are the ratios of the corresponding upper-panel chords;
they match the endpoints by construction and are not independent fits to $\hc_{\rm LS}^{(1)}$. The large enhancement of each collision moment is
substantially reduced in their ratio; the calculation does not by itself establish the
asymptotic endpoint expansion or an exact cancellation.}
\label{fig:log}
\end{figure}

The numerical calculation also resolves the generation and relaxation moments separately. The results are presented in Fig.~\ref{fig:log}. 
Over the sampled temperature interval, each separate moment is close to linear in
$L=\ln(p_Fv_F/T)$. For each moment we draw the chord through its two temperature
endpoints. The ratio of these chords follows the normalized one-mode coefficient
$\hc_{\rm LS}^{(1)}(T)/\hc_{\rm LS}^{(1)}(T_0)$, with $T_0=0.02\,\eF$,
to within $\lesssim3\times10^{-3}$ at the interior points; agreement at the endpoints
is by construction. If both leading
coefficients in Eq.~\eqref{eq:collinear-expansion} are nonzero, then generically
\begin{equation}
 \frac{\mathcal A}{\mathcal B}=\frac{a_1}{b_1}
 \left[1+\frac{a_0/a_1-b_0/b_1}{L}+O(L^{-2})\right].
\end{equation}
The data therefore support a correlated collinear enhancement of generation and relaxation, but
do not distinguish this possible $1/L$ correction from regular finite-temperature corrections,
nor do they establish the exact endpoint coefficients. No $T\to0$ intercept is inferred.

\subsection{The dc limit}
\label{sec:dc-window}

In the dc limit Eq.~\eqref{eq:effective-velocity} reduces to
\begin{equation}
v_{{\rm dc},x}=v_x-\gamma_{\rm tr}(\varepsilon)p_x\frac{\mathcal V}{\mathcal D},
\qquad \langle\Phi_0|v_{{\rm dc},x}\rangle=0,
\label{eq:dc-velocity}
\end{equation}
and the interparticle contribution is
\begin{equation}
\sigma_{xy}^{\rm dc}\simeq
-2e^2\langle v_{{\rm dc},x}|\hat R_{\rm ee}|g_y\rangle.
\label{eq:dc-sigma}
\end{equation}
Both $\gamma_{\rm tr}$ and $\mathcal D$ are proportional to $\mathcal W$, so the effective drive is independent
of its overall magnitude. Its shape still depends on the energy dependence of disorder scattering.
The numerical electron--electron matrix and generation vector are unchanged; only the drive
vector in Eq.~\eqref{eq:sigma-shift-multimode} is replaced by
$D_a^{\rm dc}=\langle\bar\chi_a\cos\vartheta|v_{{\rm dc},x}\rangle$.

For $T\ll\eF-|m|$, use $v_x=(v^2/\varepsilon)p_x$ and
$\mathcal V/\mathcal D\simeq v^2/[\eF \gamma_{\rm tr}(\eF)]$. Expansion in $\xi=\varepsilon-\eF$ gives
\begin{equation}
\begin{aligned}
v_{{\rm dc},x}&\simeq-\frac{v^2p_F}{\eF^2}
\left[1+\eF\frac{\gamma'_{\rm tr}(\eF)}{\gamma_{\rm tr}(\eF)}\right]\xi\cos\vartheta,\\
u_{\mathsf Q}\cos\vartheta&\simeq-\frac{v^2p_F}{\eF^2}\xi\cos\vartheta.
\end{aligned}
\label{eq:dc-thermal-drive}
\end{equation}
The derivative is with respect to scattering energy at fixed band parameters and fixed disorder
correlator. The exact momentum projections must be retained for finite-$T$ calculations.
For the white-noise rate~\eqref{eq:dis-rate}, the leading thermal result is
\begin{equation}
\begin{aligned}
\sigma_{xy}^{\rm dc}&\simeq R_{\rm dc}(r)\,\sigma_{xy}^{\rm cl},\\
R_{\rm dc}(r)&=1+\eF\frac{\gamma'_{\rm tr}(\eF)}{\gamma_{\rm tr}(\eF)}
=\frac{2}{1+3r^2},\qquad r=\frac{m}{\eF}.
\end{aligned}
\label{eq:dc-factor}
\end{equation}
This is one multiplier, not its square: the velocity vertex changes while the generation
vertex does not. By Eq.~\eqref{eq:sigma-2x}, shift accumulation and anomalous distribution
each acquire the same multiplier and each remains half of the total interparticle contribution.
Its positivity preserves the clean Hall sign, and its evenness in $m$ preserves the oddness
of the conductivity. The effect is enhanced for $|r|<1/\sqrt3$ and suppressed for larger
$|r|$ within the degenerate regime. At $m=0$ both conductivities vanish; the ratio denotes
the limit from nonzero mass.

\begin{figure}[tbp]
\centering
\begin{tikzpicture}
\begin{axis}[width=0.98\columnwidth,height=0.73\columnwidth,
xlabel={$m/\eF$},ylabel={$\hc$},xmin=0,xmax=0.95,ymin=0,ymax=0.26,
legend style={at={(0.5,1.03)},anchor=south,legend columns=2,draw=none,fill=none,
font=\footnotesize,/tikz/column sep=3pt,/tikz/row sep=1pt},legend cell align=left,
scaled y ticks=false,yticklabel style={/pgf/number format/fixed,/pgf/number format/precision=2},
grid=major,grid style={gray!20}]
\addplot[blue!70!black,thick,mark=triangle*,mark size=1.7pt] coordinates {
(0.1,0.028719410)(0.2,0.057476887)(0.3,0.086317351)
(0.4,0.115282959)(0.5,0.144438940)(0.6,0.173845419)
(0.7,0.203468115)(0.8,0.231952435)};
\addlegendentry{Clean}
\addplot[red!70!black,thick,mark=*,mark size=1.7pt] coordinates {
(0.1,0.055585631)(0.2,0.102360103)(0.3,0.135671481)
(0.4,0.155635904)(0.5,0.165097057)(0.6,0.167419985)
(0.7,0.165355148)(0.8,0.160117976)};
\addlegendentry{dc}
\addplot[black!60,dashed,domain=0.1:0.8,samples=80] {0.58*x/(1+3*x^2)};
\addlegendentry{dc estimate}
\addplot[only marks,orange!80!black,mark=diamond*,mark size=2.2pt] coordinates {(0.9,0.234172629)};
\addlegendentry{Clean, $r=0.9$}
\addplot[only marks,red!70!black,mark=diamond,mark size=2.2pt,thick] coordinates {(0.9,0.139418998)};
\addlegendentry{dc, $r=0.9$}
\end{axis}
\end{tikzpicture}
\caption{Finite-temperature interparticle shift coefficients in the clean and dc limits,
at $T/\eF=0.02$. The clean values are identical to the finite-$T$ series in
Fig.~\ref{fig:h}; the dc values use the same collision matrices with the disorder-modified
effective drive. Connecting segments guide the eye only over $0.1\leq r\leq0.8$.
At $r=0.9$, where $\eF-|m|=5T$, separate diamonds show the band-edge audit:
filled orange for the clean limit, open red for dc. These points retain the finite-temperature
band-edge departure from the nearly linear clean series; no interpolation to them is implied.
The dashed dc estimate, $0.58r/(1+3r^2)$, combines the empirical clean fit
$\hc\simeq0.29r$ with the leading thermal multiplier over the fit's mass interval. It is not an independent determination of a strict
$T\to0$ coefficient.}
\label{fig:dc-comparison}
\end{figure}
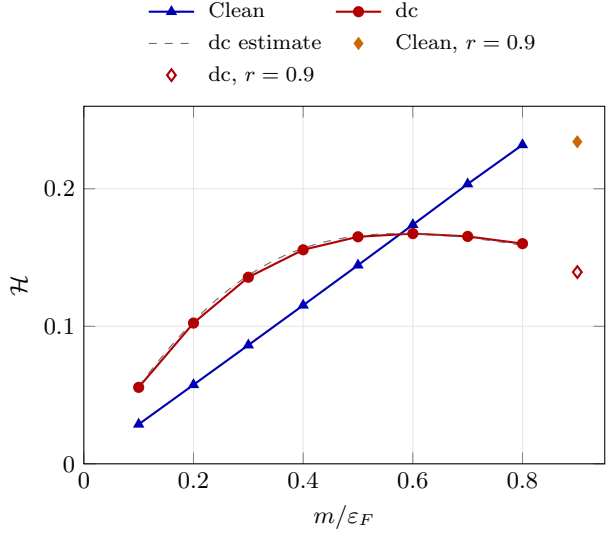

At finite temperature we evaluate Eq.~\eqref{eq:dc-velocity} directly, using the moments in
Appendix~\ref{app:disorder-projection}, instead of imposing Eq.~\eqref{eq:dc-factor}.
The result is written as $\sigma_{xy}^{\rm dc}=(e^2/\hbar)t^2\hc_{\rm dc}(r;t)$, with
$t=T/\eF$. Fig.~\ref{fig:dc-comparison} compares it with
the clean coefficient $\hc_{\rm cl}\equiv\hc$.

\subsection{Frequency crossover and scope of the dc result}
\label{sec:crossover}

The clean and dc limits follow from a common frequency-dependent response. At any temperature within the conduction-band
kinetic model, Eq.~\eqref{eq:effective-velocity} gives exactly
\begin{equation}
v_{{\rm eff},\omega}
=u_{\mathsf Q}\cos\vartheta+\frac{\Gamma}{\Gamma-i\omega}
\big(v_{{\rm dc},x}-u_{\mathsf Q}\cos\vartheta\big).
\label{eq:drive-interpolation}
\end{equation}
Substitution in the leading hydrodynamic bilinear~\eqref{eq:hydro-bilinear} yields
\begin{equation}
\sigma_{xy}^{\rm shift}(\omega)\simeq\sigma_{xy}^{\rm cl}
+\frac{\Gamma}{\Gamma-i\omega}
\big(\sigma_{xy}^{\rm dc}-\sigma_{xy}^{\rm cl}\big).
\label{eq:frequency-interpolation}
\end{equation}
This interpolation requires $\Gamma,\omega\ll\gamma_{\rm ee}$, but does not require a
Sommerfeld expansion: its endpoints may be the full finite-$T$ Galerkin coefficients.
The exact reduced kinetic operator in Appendix~\ref{app:disorder-projection} supplies the
relaxation corrections outside this leading approximation.
With the additional low-temperature expansion,
\begin{equation}
\frac{\sigma_{xy}^{\rm shift}(\omega)}{\sigma_{xy}^{\rm cl}}
\simeq1+\frac{1-3r^2}{1+3r^2}\frac{\Gamma}{\Gamma-i\omega}.
\label{eq:frequency-lowT}
\end{equation}
For the $e^{-i\omega t}$ convention the imaginary part of the second term has the sign of
$(1-3r^2)\omega$. Within this low-temperature result, no additional sign change occurs between the two real limiting coefficients.

The magnitude of the disorder drops out of the leading dc coefficient, but its energy dependence
does not. An energy-independent $\hat D$ would have $d=0$ and give identical endpoints;
the microscopic rate~\eqref{eq:dis-rate} generally does not. The Born factor $\lambda^2$
cancels between the common generation and relaxation operators in both limits, while the
explicit $T^2$ prefactor and any residual collinear dependence remain. At fixed disorder,
lowering $T$ eventually violates $\Gamma\ll\gamma_{\rm ee}$; the low-temperature statements require
maintaining this hierarchy, rather than taking a fixed-disorder limit all the way to $T=0$.

The calculated term can coexist with intrinsic, impurity-induced, and interaction-renormalization
backgrounds. Their responses must be treated with the same interacting kinetic operator before
a total measured Hall conductivity is assembled. In particular, the disorder-only result of
Ref.~\cite{ado15} cannot simply be added unchanged. A finite channel requires its own
boundary-value problem; no claim is made that wall relaxation is equivalent to the white-noise
bulk kernel studied here. 

\section{Summary, conclusions, and outlook}
\label{sec:summary}

We have calculated the interparticle coordinate-shift Hall response of a two-dimensional
massive Dirac Fermi liquid with a weak contact interaction and weak scalar Gaussian
white-noise disorder. The finite-frequency kinetic equation covers both the clean limit, $\Gamma\ll\omega\ll\gamma_{\rm ee}$, and the dc limit,
$\omega\ll\Gamma\ll\gamma_{\rm ee}$. Writing $r=m/\eF$ and $t=T/\eF$, their responses take
the form
\begin{equation}
\sigma^a_{xy}\simeq\frac{e^2}{\hbar}t^2\hc_a(r;t),
\qquad a\in\{\mathrm{cl},\mathrm{dc}\}.
\label{eq:final-shift}
\end{equation}
The clean coefficient $\hc_{\rm cl}\equiv\hc$ is approximately $0.29r$ at $t=0.02$ over
$0.1\leq|r|\leq0.8$. In the dc limit, the leading low-temperature result is
$\hc_{\rm dc}\simeq 2\hc_{\rm cl}/(1+3r^2)$ [Eq.~\eqref{eq:dc-factor}]. This enhances the
shift contribution for $|r|<1/\sqrt3$ and suppresses it for larger $|r|$ within the degenerate
regime. Fig.~\ref{fig:dc-comparison} gives both coefficients
with the full thermal width retained; Eq.~\eqref{eq:frequency-interpolation} describes their
leading hydrodynamic frequency crossover.

The different limits reflect the coupling of conserved momentum to relaxing distribution
modes by energy-dependent impurity scattering. This coupling is weak, but the driven momentum
response grows as $1/\Gamma$ in dc, leaving a finite correction when the disorder strength
is taken to zero after the frequency. Shift accumulation and anomalous distribution remain
equal in both windows. Both limiting results are odd in $m$, oppose the intrinsic Hall
response in the computed regime, and are independent of interaction strength at leading Born
order. Their explicit thermal suppression comes from the off-shell coordinate shift and the
projected effective drive. For a parabolic dispersion the clean velocity residue vanishes,
as does the projected clean coordinate-shift charge response. The asymptotic collinear
dependence of the coefficients is not established by the finite-window numerical data.
At fixed disorder the hierarchy $\Gamma\ll\gamma_{\rm ee}$ eventually fails as $T$ is
lowered, so these results do not describe the fixed-disorder limit all the way to $T=0$.

Just as impurity scattering changes the interparticle shift response in dc, electron--electron
collisions generally modify the impurity-induced AHE. Impurity side-jump and skew-scattering currents depend on
distributions obtained from the combined kinetic operator, and their disorder-only values
cannot in general be carried unchanged into the hydrodynamic regime. Such modifications
have been studied in channel geometries \cite{zohrabyan23}, but are not calculated for the
present massive Dirac model. In particular, a complete treatment for Gaussian disorder must
retain the coherent impurity-pair scattering responsible for part of the skew response
\cite{ado15}. The present dc coefficient includes disorder in the relaxation operator, but
not these additional impurity Hall vertices; it is therefore an interparticle contribution,
not the total Hall conductivity. No common multiplicative correction to all mechanisms is
implied by Eq.~\eqref{eq:dc-factor}.

The fixed-band intrinsic response provides a useful thermal comparison. Let $\eF$ be the
zero-temperature chemical potential and hold $m,v$ fixed. For $T\ll\eF-|m|$, including the
filled valence band gives
\begin{equation}
\sigma^{\rm int}_{xy}(T)=-\frac{e^2}{4\pi\hbar}\frac{m}{\eF}
+\delta\sigma^{\rm int}_{xy}(T),
\end{equation}
with the leading Sommerfeld corrections
\begin{equation}
\begin{aligned}
\delta\sigma^{\rm int}_{xy}(T)\big|_{\mu,n}
&=-c_{\mu,n}\frac{e^2}{\hbar}\frac{m}{\eF}\left(\frac{T}{\eF}\right)^2,\\
c_\mu&=\frac{\pi}{12},\qquad c_n=\frac{\pi}{8}.
\end{aligned}
\label{eq:intT}
\end{equation}
The fixed-density coefficient includes
$\mu(T)=\eF-\pi^2T^2/(6\eF)+O(T^4/\eF^3)$. These coefficients are unchanged by the
order of the $\omega,\Gamma$ limits at the semiclassical order retained here: the intrinsic
current is the anomalous velocity, already linear in the field, averaged over the
equilibrium distribution. It does not acquire the disorder-dependent effective drive of the
collisional shift term. Interaction renormalization of the intrinsic response, including the
first-order correction of Ref.~\cite{dumitriu24}, and disorder corrections to the band itself
are excluded from this fixed-band comparison.

The intrinsic and interparticle thermal terms thus have opposite signs in both windows,
but their balance depends on the order of limits. Using the empirical clean fit in the
computed temperature window, the coefficients multiplying
$(e^2/\hbar)r t^2$ in their sum are approximately
\[
0.29-c_{\mu,n}\quad\text{(clean)},\qquad
\frac{0.58}{1+3r^2}-c_{\mu,n}\quad\text{(dc)}.
\]
The dc estimate additionally uses the leading thermal multiplier.
To leading thermal order away from the band edge, the fixed-density shift of $\mu$ changes
the collisional term only beyond its explicit $T^2$ order. In the clean limit, the numerical
differences remain $0.29-\pi/12\simeq0.028$ at fixed $\mu$ and
$0.29-\pi/8\simeq-0.103$ at fixed density. In dc the corresponding balance is mass dependent,
so the clean near cancellation is not a general dc result. These comparisons neither establish
an asymptotic cancellation as $T\to0$ nor predict the total temperature-dependent Hall response once the
omitted impurity and interaction-renormalization contributions are included.

Frequency-dependent measurements could probe the $\Gamma$-scale crossover in
Eq.~\eqref{eq:frequency-lowT}, whose relative amplitude $(1-3r^2)/(1+3r^2)$ is absent
from the fixed-band intrinsic term at the order retained here. Impurity Hall mechanisms
can also disperse on this scale, however, so identifying the interparticle contribution
requires analyzing these backgrounds together.

Interparticle skew scattering provides another possible anomalous transport mechanism. It
has been considered for spin and valley transport \cite{glazov22} and for Hall Coulomb drag
\cite{gutman24}; impurity- and phonon-induced skew scattering across the hydrodynamic crossover
was considered in Ref.~\cite{zohrabyan23}. For the single-species contact model,
Eq.~\eqref{eq:factor} makes the two-particle ladder amplitude separable, so its probability is
reciprocal to all orders within that ladder series. At leading $O(\lambda^3)$, the crossed
contribution also vanishes when all participating states lie on a common energy shell:
in a suitable gauge every contact vertex on that shell is real, so its imaginary interference
product is zero. These restricted cancellations do not establish the absence of every
finite-temperature crossed or higher-order contribution. A full treatment of interparticle
skew scattering, including long-range interactions, is left for future work.

Other open questions include a controlled asymptotic evaluation of the collinear collision
moments and their ratio (Sec.~\ref{sec:logs}), dynamically screened Coulomb interactions,
and a quantitative comparison with the first-order intrinsic correction of
Ref.~\cite{dumitriu24}. Multivalley systems require the distinguishable-species collision
channels to be included; time-reversed valleys cancel the charge Hall response at equal
populations and naturally support valley or spin Hall responses. Finally, a finite-channel
calculation could combine the present bulk sources with the hydrodynamic framework of
Ref.~\cite{zohrabyan23}. Hall viscosity and anomalous thermal transport require separate
conserving derivations, including the relevant magnetization contributions.

\begin{acknowledgments}
We thank Leonid Golub, Elio K\"onig, and Songci Li for their interest in this problem and numerous useful discussions.
The work of A. L. was supported by NSF Grant No. DMR-2452658 and H. I. Romnes Faculty Fellowship provided by the University of Wisconsin-Madison Office of the Vice Chancellor for Research and Graduate Education with funding from the Wisconsin Alumni Research Foundation. This work was performed in part during the workshop program "Emerging New Phases in Quantum Materials: The Disordered, the Strange and the Topological" at the Aspen Center for Physics, which is supported by National Science Foundation grant PHY-2210452. The authors acknowledge the use of Claude (Anthropic) and ChatGPT (OpenAI) in manuscript preparation, which includes numerical calculations, graphics, and symbolic verification of analytical results. The authors conceived the project and carried out the research.
\end{acknowledgments}

\appendix

\section{Wave-packet derivation of the single-particle side jump}
\label{app:sj}

Consider a packet built from the states of band $n$,
\begin{equation}
|\Psi_{\rm in}\rangle = \int \! d\kk\, a(\kk-\kk_0)\,e^{i\kk\cdot\rr}\,|u_{\kk}\rangle ,
\end{equation}
with a sharply peaked envelope $a(\kk-\kk_0)$. The packet center follows from the stationary phase of the superposition
and contains, besides the trivial group motion, the Berry-connection term: for an envelope $a(\kk) = |a|e^{i\eta(\kk)}$, the center is
$\rr_c = -\nabla_{\kk}\eta|_{\kk_0} + \bm A(\kk_0)$. The outgoing wave in the direction of
$\kk'$ carries the amplitude $\propto V_{\kk'\kk}\,a(\kk-\kk_0)$, i.e., its envelope inherits the
phase of the matrix element. Repeating the stationary-phase construction for the outgoing
packet, the phase now consists of three pieces: the envelope phase $\eta$, the Bloch phase of
$|u_{\kk'}\rangle$, and $\arg V_{\kk'\kk}$. Taking the difference of the outgoing and incoming
packet centers at the moment of scattering gives the coordinate shift \eqref{eq:sj} of the main
text.

For a scalar scatterer its Fourier phase cancels exactly as explained in
Sec.~\ref{sec:sj}. For small momentum transfer $\qq=\kk'-\kk$, using
$\langle u_{\kk+\qq}|u_{\kk}\rangle = 1 + i\qq\cdot\bm A(\kk) +
O(q^2)$, so that $\arg\langle u_{\kk'}|u_{\kk}\rangle = (\kk'-\kk)\cdot \bm A\big(\tfrac{\kk+\kk'}
{2}\big) + O(q^3)$, one finds
\begin{equation}
\begin{aligned}
&\big(\nabla_{\kk'}+\nabla_{\kk}\big)_i \arg\langle u_{\kk'}|u_{\kk}\rangle = q_j\,\partial_i A_j + O(q^3),\\
&\big[\bm A(\kk+\qq)-\bm A(\kk)\big]_i = q_j\,\partial_j A_i + O(q^3),
\end{aligned}
\end{equation}
with derivatives evaluated at the midpoint $(\kk+\kk')/2$, and hence
\begin{equation}
\begin{aligned}
\delta r_i&=q_j\big(\partial_j A_i-\partial_i A_j\big)+O(q^3)\\
&=-\epsilon_{ij}q_j\,\Omega\!\left(\frac{\kk+\kk'}{2}\right)+O(q^3).
\end{aligned}
\end{equation}
This is Eq.~\eqref{eq:sjsmallq} of the main text (both lines of the preceding equation are
accurate to $O(q^3)$ when referred to the midpoint, the even orders cancelling by symmetry).

\section{The two-particle shift for 2D massive Dirac fermions}
\label{sec:dirac}

\subsection{Model}

The minimal band structure exhibiting the physics of collisional shifts must break time-reversal
(or inversion, per valley), possess Berry curvature, and be nonparabolic; the 2D massive
Dirac Hamiltonian is the minimal such model:
\begin{equation}
\begin{aligned}
&H = v(k_x\sigma_x + k_y\sigma_y) + m\sigma_z,
\qquad
\varepsilon_{\kk} = \sqrt{v^2k^2+m^2},\\
&|u_{\kk}\rangle = \begin{pmatrix} \cos\frac{\theta_{\kk}}{2}\\[2pt]
\sin\frac{\theta_{\kk}}{2}\,e^{i\vartheta_{\kk}}\end{pmatrix},
\qquad \cos\theta_{\kk} = \frac{m}{\varepsilon_{\kk}} .
\end{aligned}
\label{eq:model}
\end{equation}
Here $\vartheta_{\kk}$ is the momentum azimuth and $\theta_{\kk}$ is the Bloch-sphere
polar angle. Throughout, $\hbar=k_{\rm B}=1$ in intermediate expressions. The Fermi level lies in the conduction band,
$\eF = \sqrt{v^2p_F^2+m^2}$,
$v_F = v^2p_F/\eF$, $\nu_F = \eF/2\pi v^2$, and the interaction is a weak short-range (contact)
repulsion $U = \lambda$, with dimensionless strength $\nu_F\lambda\ll1$; a single spinless cone is
considered. The Sommerfeld expansion assumes $T\ll\eF-|m|$; finite-temperature numerical
results near the band edge are identified separately. Except when explicitly discussing fixed
density, the numerical calculation holds $\mu=\eF$ fixed. The quantitative results in Sec.~\ref{sec:conductivity} refer to this nondegenerate species: for
distinguishable flavors (spin or valley) the direct and exchange terms of the pair amplitude
enter with different weights, the factorization of Appendix~\ref{app:pair-factorization} does not apply to
the interflavor channel, and the collision moments must be recomputed; in a two-valley
realization with opposite $m$ the charge response moreover requires valley imbalance, exactly as
the intrinsic AHE. The band geometry is encoded in
\begin{equation}
\bm A(\kk) = -\frac{1}{2}\Big(1-\frac{m}{\varepsilon_{\kk}}\Big)\frac{\zhat\times\hat\kk}{k},
\qquad
\Omega(\kk) = -\frac{m v^2}{2\varepsilon_{\kk}^3} .
\label{eq:berry}
\end{equation}

\subsection{Factorization of the pair amplitude}
\label{app:pair-factorization}

Introduce the stereographic variable of the Bloch sphere,
\begin{equation}
\begin{aligned}
&z_{\kk} \equiv \tan\frac{\theta_{\kk}}{2}\,e^{i\vartheta_{\kk}}
= \frac{vk}{\varepsilon_{\kk}+m}\,e^{i\vartheta_{\kk}},\\
&\langle u_{\kk'}|u_{\kk}\rangle
= \cos\frac{\theta_{\kk'}}{2}\cos\frac{\theta_{\kk}}{2}\,\big(1+\bar z_{\kk'} z_{\kk}\big).
\end{aligned}
\label{eq:z}
\end{equation}
Substituting into the antisymmetrized amplitude \eqref{eq:V2} with $U = \lambda$, the constant and quartic terms
in the $z$'s cancel between direct and exchange (Appendix~\ref{app:stereographic}), and
the amplitude factorizes exactly into initial- and final-pair factors:
\begin{equation}
\frac{V}{\lambda N} = \big(\bar z_{\pp_1}-\bar z_{\pp_2}\big)\big(z_{\kk_1}-z_{\kk_2}\big),
\qquad N = \prod_i\cos\frac{\theta_i}{2} .
\label{eq:factor}
\end{equation}
In this northern-gauge patch $N$ is real and nonnegative. For $m<0$ the patch is singular only at
the measure-zero point $k=0$; switching gauge there leaves Eqs.~\eqref{eq:argV}--\eqref{eq:W2}
and every gauge-invariant result unchanged.
The factorization implies
\begin{subequations}
\begin{equation}
\arg V = \arg\big(z_{\kk_1}-z_{\kk_2}\big) - \arg\big(z_{\pp_1}-z_{\pp_2}\big),
\label{eq:argV}
\end{equation}
\begin{equation}
|V|^2 = \frac{\lambda^2}{4}\,\big(1-\bm n_{\kk_1}\!\cdot\bm n_{\kk_2}\big)
\big(1-\bm n_{\pp_1}\!\cdot\bm n_{\pp_2}\big),
\label{eq:W2}
\end{equation}
\end{subequations}
where $\bm n_{\kk}=(v\kk,m)/\varepsilon_{\kk}$.
At $m=0$, $|z_{\kk}| = 1$ for all nonzero $\kk$, so that
two co-directed massless particles have $z_{\kk_1} = z_{\kk_2}$ and cannot interact locally:
the Pauli principle acting through the spinor structure. Equation \eqref{eq:W2} is the
statement that the scattering probability vanishes quadratically, as $|\bm n_a-\bm n_b|^2$, when
the Bloch vectors of the two partners of a pair approach each other, in the initial and in the
final state; this is what will make the collision moments of the shift converge without any
external cutoff.

\subsection{Closed form of the shift}
\label{sec:closed}

Because $\arg V$ separates into initial- and final-pair pieces [Eq.~\eqref{eq:argV}], so does
the shift \eqref{eq:master}: it is the difference of one and the same pair function evaluated on
the final and on the initial momenta. The same structure, and a closed form for the pair
function, are available for 3D Weyl fermions, $H_{\rm W} = v\,\bm K\cdot\bm\sigma$, with the
same local interaction $U = \lambda$. The factorization \eqref{eq:factor} uses only the
Bloch-sphere form of the spinors and holds there verbatim, with $z$ the stereographic coordinate
of the direction of $\bm K$, and Eq.~\eqref{eq:master} evaluates to \cite{pesin18}

\begin{equation}
\begin{aligned}
&\dr^{\pp_1\pp_2}_{\kk_1\kk_2} = \bm P_{\rm W}(\pp_1,\pp_2)-\bm P_{\rm W}(\kk_1,\kk_2),\\
&\bm P_{\rm W}(\bm a,\bm b) = \frac{v}{2}\Big(\frac{1}{\varepsilon_b}-\frac{1}{\varepsilon_a}\Big)
\frac{\bm e_a\times\bm e_b}{1-\bm e_a\!\cdot\bm e_b},
\qquad \bm e_{\bm a} = \frac{v\bm a}{\varepsilon_a},
\end{aligned}
\label{eq:weyl}
\end{equation}
where all momenta are three-dimensional, $\varepsilon_{\bm a} = va$, and $\bm e_{\bm a}$ is the
Bloch vector of the state $\bm a$. Two features of Eq.~\eqref{eq:weyl} anticipate everything
that follows: each pair function contains the difference of the partners' inverse energies, and it
has a pole for co-directed Weyl partners of unequal energy ($\bm e_a\cdot\bm e_b\to1$ with
$\varepsilon_a\neq\varepsilon_b$); the fate of this pole in the 2D massive model is discussed in
item (c) below.

The 2D massive Dirac model inherits Eq.~\eqref{eq:weyl} through a Bloch-sphere embedding. The
Hamiltonian \eqref{eq:model} is the Weyl Hamiltonian evaluated on the plane $K_z = m/v$ of
three-dimensional momentum space, $H = v\,\bm K_{\kk}\cdot\bm\sigma$ with
$\bm K_{\kk} \equiv (\kk,\,m/v)$: the spinors coincide, $\varepsilon_{\bm K_{\kk}} =
\varepsilon_{\kk}$, and the Bloch vector $\bm e_{\bm K_{\kk}} = (v\kk,\,m)/\varepsilon_{\kk} =
\bm n_{\kk}$ is precisely the unit vector of Eq.~\eqref{eq:W2}. Every ingredient of
Eq.~\eqref{eq:master} for the 2D problem is then the in-plane component of its Weyl counterpart
evaluated on this plane. The connection \eqref{eq:berry} is the in-plane part of the Weyl Berry
connection, because only in-plane derivatives of the common spinor enter it. A contact
interaction carries no form factor, so the Weyl pair amplitude with all four momenta on the
plane is the 2D amplitude \eqref{eq:V2} itself; and restricting all four momenta to the plane is
consistent with momentum conservation ($K_{1z}+K_{2z} = 2m/v = K_{3z}+K_{4z}$ is then
automatically satisfied), so that the 2D kinematics is a consistent section of the 3D one and the
in-plane gradients of $\arg V$ coincide. Hence the 2D shift
is the in-plane projection of Eq.~\eqref{eq:weyl} with $\bm e\to\bm n$; the $z$ component of the
Weyl shift has no 2D counterpart and is discarded. The projection is elementary: for $\bm K_{\bm a} = (\bm a,\,m/v)$ and
$\bm K_{\bm b} = (\bm b,\,m/v)$,
\begin{equation}
\begin{aligned}
&1-\bm n_a\!\cdot\bm n_b = \frac{\varepsilon_a\varepsilon_b - m^2 - v^2\,\bm a\cdot\bm b}
{\varepsilon_a\varepsilon_b},\\
&\bm n_a\times\bm n_b = \frac{mv\,(\bm a-\bm b)\times\zhat + v^2\,\bm a\times\bm b}
{\varepsilon_a\varepsilon_b},
\end{aligned}
\label{eq:embed}
\end{equation}
where $\bm a\times\bm b\parallel\zhat$ is the out-of-plane part. Substituting into
Eq.~\eqref{eq:weyl} and keeping the in-plane part,
\begin{equation}
\begin{aligned}
\dr^{\pp_1\pp_2}_{\kk_1\kk_2} &= \bm P(\pp_1,\pp_2)-\bm P(\kk_1,\kk_2),\\
\bm P(\bm a,\bm b) &= \big[\bm P_{\rm W}(\bm K_{\bm a},\bm K_{\bm b})\big]_\parallel\\
&= \frac{mv^2}{2}
\Big(\frac{1}{\varepsilon_b}-\frac{1}{\varepsilon_a}\Big)
\frac{(\bm a-\bm b)\times\zhat}{\varepsilon_a\varepsilon_b - m^2 - v^2\,\bm a\cdot\bm b} .
\end{aligned}
\label{eq:P}
\end{equation}
The direct evaluation of Eq.~\eqref{eq:master} in two dimensions, carried out in the
stereographic variables in Appendix~\ref{app:stereographic}, reproduces the pair function
\eqref{eq:P} identically [Eq.~\eqref{eq:R}] and thus confirms the embedding; the numerical implementation uses the algebraically regularized Cartesian product
$|V|^2\dr$ described below to evaluate the collision moments in
Sec.~\ref{sec:hplot}. In the form \eqref{eq:P} the following properties are manifest:

\emph{(a) Chirality.} $\dr$ is odd in $m$ and vanishes at $m=0$ wherever the scattering amplitude is nonzero.
At zero-amplitude configurations the shift itself need not be defined, but the
amplitude-weighted shift that enters transport vanishes. Geometrically, the
mass is the offset of the embedding plane from the origin: at $m=0$ all Bloch vectors
$\bm n_{\kk}$ lie in the equatorial plane of the Bloch sphere, $\bm n_a\times\bm n_b$ is purely
out of plane, and the in-plane projection of the Weyl shift vanishes, as also follows directly
from Eq.~\eqref{eq:R}.

\emph{(b) Energy-difference structure.} Each pair term is proportional to
$(1/\varepsilon_b - 1/\varepsilon_a)$: the shift vanishes whenever the two partners of each
pair have equal energies. On a single Fermi surface this is the generic situation, and it is the
central kinematic fact of the problem. In particular, for strictly elastic kinematics
(all four momenta on the Fermi surface) $\dr \equiv 0$ pointwise. For nonzero total momentum,
two-dimensional energy and momentum conservation fix the unordered equal-energy pair up to
exchange. The exceptional head-on manifold of zero total momentum permits arbitrary rotations of
an antipodal pair, but Eq.~\eqref{eq:P} still makes its coordinate shift vanish.

\emph{(c) Collinear structure and boundedness.} By Eq.~\eqref{eq:embed} the denominator of
$\bm P$ is $\varepsilon_a\varepsilon_b\,(1-\bm n_a\!\cdot\bm n_b) =
\tfrac12\varepsilon_a\varepsilon_b\,|\bm n_a-\bm n_b|^2$, which vanishes only when the two Bloch
vectors coincide. For $m\neq0$ this happens only at $\bm a = \bm b$: co-directed partners of
unequal energy have different $n_z = m/\varepsilon$, so the pole of the Weyl pair function is not
inherited by the 2D section [the plane $K_z = m/v\neq0$ contains no two distinct co-directed
three-dimensional momenta]. In the limit $\bm a\to\bm b$ the numerator
$(1/\varepsilon_b-1/\varepsilon_a)(\bm a-\bm b)\times\zhat$ is also of second order in
$\bm a-\bm b$, and $\bm P$ tends to a finite, direction-dependent limit: with
$\bm a = \bm b+\bm\delta$,
\begin{equation}
\bm P(\bm b+\bm\delta,\bm b)\;\xrightarrow[\ \delta\to0\ ]{}\;
\frac{m v^2\,(\bm b\cdot\bm\delta)\,(\bm\delta\times\zhat)}
{\varepsilon_b^3\,\big[\delta^2 - v^2(\bm b\cdot\bm\delta)^2/\varepsilon_b^2\big]},
\label{eq:Plimit}
\end{equation}
which vanishes for transverse approach and reaches $|\bm P| = v_b/|m|$, with $v_b = v^2b/\varepsilon_b$
the group velocity, for approach along $\hat{\bm b}$. The pair function is therefore bounded, and
the small-$q$ form \eqref{eq:drFS} below, with its factor $1/(1-\hat\pp_1\cdot\hat\pp_2)$, is a
nonuniform expansion valid for $|\hat\pp_1-\hat\pp_2|\gg q/p_F$ rather than the signature of a
divergence. As $m\to0$ the bound $v_b/|m|$ grows, in keeping with the pole of the massless Weyl
function, but the region in which it is approached shrinks, and in every collision moment the
Pauli factor \eqref{eq:W2} of each pair multiplies the denominator of the corresponding pair
term,
\begin{equation}
\begin{aligned}
W\,\dr\propto{}&\big(1-\bm n_{\kk_1}\!\cdot\bm n_{\kk_2}\big)
\big(1-\bm n_{\pp_1}\!\cdot\bm n_{\pp_2}\big)\\
&\times\big[\bm P(\pp_1,\pp_2)-\bm P(\kk_1,\kk_2)\big],
\end{aligned}
\label{eq:Wdr}
\end{equation}
so that $W\dr$ vanishes, rather than merely stays finite, whenever either pair becomes
degenerate. The contact model thus requires no screening cutoff (Sec.~\ref{sec:conductivity}).

\subsection{Limiting forms}
\label{sec:limits}

\emph{(i) Small momentum transfer, long-range interaction.} If the interaction is dominated by its
direct, small-$q$ part (e.g., screened Coulomb with $\varkappa\ll p_F$), the exchange term in
Eq.~\eqref{eq:V2} may be dropped, $\arg V = \arg\langle u_{\pp_1}|u_{\kk_1}\rangle + \arg\langle
u_{\pp_2}|u_{\kk_2}\rangle$, and the two-particle shift becomes the sum of two independent side
jumps \eqref{eq:sjsmallq} with transfers $\pm\qq$:
\begin{equation}
\dr = \big[\Omega(\kk_1)-\Omega(\kk_2)\big]\,\zhat\times\qq + O(q^2).
\label{eq:smallq}
\end{equation}
The difference of Berry curvatures appears because the second particle absorbs $-\qq$;
Eq.~\eqref{eq:smallq} is the long-range counterpart of the $(1/\varepsilon_b-1/\varepsilon_a)$
prefactor of Eqs.~\eqref{eq:weyl} and \eqref{eq:P} and its displayed leading term again vanishes for equal-energy partners.

\emph{(ii) Near the Fermi surface.} For the contact model with the initial pair on the Fermi surface and
small transfer $\qq$ (finals off shell by the energy transfer), expanding the prefactor of the
final-pair term of Eq.~\eqref{eq:P} gives
\begin{equation}
\dr(\pp_1\pp_2;\qq) = \frac{mv^2 q}{2\eF^3}\,
\big[(\hat\pp_1+\hat\pp_2)\cdot\hat\qq\big]\,
\frac{(\hat\pp_1-\hat\pp_2)\times\zhat}{1-\hat\pp_1\cdot\hat\pp_2} + O(q^2),
\label{eq:drFS}
\end{equation}
where $(\vv_{\pp_1}+\vv_{\pp_2})\cdot\qq$ measures precisely the energy off-shellness generated by
the transfer, and the expansion holds for $|\hat\pp_1-\hat\pp_2|\gg q/p_F$ [item (c) above]. In
the thermal window the off-shellness factor is $O(T)$: the shift active in transport is
linear in the thermal off-shellness of the colliding pair. The transport results of Sec.~\ref{sec:conductivity} follow from
this fact combined with the conservation laws established in Sec.~\ref{sec:conservation}.

\section{Stereographic factorization and the two representations of the shift}
\label{app:stereographic}

In the stereographic variables \eqref{eq:z}, substitution into the antisymmetrized amplitude
\eqref{eq:V2} with $U = \lambda$ gives
\begin{align}
\frac{V}{\lambda N}
&=(1+\bar z_{\pp_1}z_{\kk_1})(1+\bar z_{\pp_2}z_{\kk_2})
\nonumber\\
&\quad-(1+\bar z_{\pp_1}z_{\kk_2})(1+\bar z_{\pp_2}z_{\kk_1})
\nonumber\\
&=\bar z_{\pp_1}z_{\kk_1}+\bar z_{\pp_2}z_{\kk_2}
-\bar z_{\pp_1}z_{\kk_2}-\bar z_{\pp_2}z_{\kk_1}
\nonumber\\
&=\big(\bar z_{\pp_1}-\bar z_{\pp_2}\big)
\big(z_{\kk_1}-z_{\kk_2}\big),
\end{align}
i.e., the constant and quartic terms cancel between direct and exchange, giving
Eq.~\eqref{eq:factor} of Appendix~\ref{sec:dirac}. The consequence \eqref{eq:W2} follows from the identity
$\cos^2\frac{\theta_1}{2}\cos^2\frac{\theta_2}{2}|z_1-z_2|^2 = 1-|\langle u_1|u_2\rangle|^2
= \tfrac12(1-\bm n_1\cdot\bm n_2)$.

Because $\arg V$ separates [Eq.~\eqref{eq:argV}], so does the shift \eqref{eq:master}:
\begin{equation}
\begin{aligned}
&\dr^{\pp_1\pp_2}_{\kk_1\kk_2} = \bm R(\pp_1,\pp_2) - \bm R(\kk_1,\kk_2),\\
&\bm R(\kk_1,\kk_2) = \bm A(\kk_1)+\bm A(\kk_2)\\
&\hspace{2.4cm}+\big(\nabla_{\kk_1}+\nabla_{\kk_2}\big)
\arg\big(z_{\kk_1}-z_{\kk_2}\big).
\end{aligned}
\label{eq:Rdef}
\end{equation}
The gradients are elementary once one notes that $|z_{\kk}|=vk/(\varepsilon+m)$, so
\begin{equation}
\begin{aligned}
&\partial_k\ln|z_{\kk}| = \frac{m}{\varepsilon k},
\qquad
\nabla_{\kk}\vartheta_{\kk} = \frac{\zhat\times\hat\kk}{k}\\
&\quad\Longrightarrow\quad
\nabla_{\kk} z_{\kk} = \frac{z_{\kk}}{k}\Big(\frac{m}{\varepsilon_{\kk}}\,\hat\kk
+ i\,\zhat\times\hat\kk\Big).
\end{aligned}
\end{equation}
Using $\nabla_{\kk_1}\arg(z_1-z_2) = {\rm Im}\,[\nabla_{\kk_1}z_1/(z_1-z_2)]$
and defining $\Delta\vartheta=\vartheta_2-\vartheta_1$, introduce
\begin{equation}
\begin{aligned}
&\zeta \equiv \frac{z_1}{z_1-z_2} = \frac{1}{1-\rho e^{i\Delta\vartheta}},
\qquad \rho = \frac{|z_2|}{|z_1|},\\
&{\rm Re}\,\zeta = \frac{1-\rho\cos(\Delta\vartheta)}{1-2\rho\cos(\Delta\vartheta)+\rho^2},\\
&{\rm Im}\,\zeta = \frac{\rho\sin(\Delta\vartheta)}{1-2\rho\cos(\Delta\vartheta)+\rho^2},
\end{aligned}
\end{equation}
one obtains, with $s_i = \sin^2(\theta_i/2) = \tfrac12(1-m/\varepsilon_i)$,
\begin{equation}
\begin{aligned}
\bm R(\kk_1,\kk_2)
&= \frac{{\rm Re}\,\zeta - s_1}{k_1}\,\zhat\times\hat\kk_1
+ \frac{1-{\rm Re}\,\zeta - s_2}{k_2}\,\zhat\times\hat\kk_2\\
&\quad+ m\,{\rm Im}\,\zeta\;\Big(\frac{\hat\kk_1}{\varepsilon_1 k_1}
- \frac{\hat\kk_2}{\varepsilon_2 k_2}\Big).
\end{aligned}
\label{eq:R}
\end{equation}
As a direct check, at $m=0$ one has $|z_1|=|z_2|=1$, $\rho=1$,
${\rm Re}\,\zeta=s_1=s_2=1/2$, and the last line is proportional to $m$; hence
$\bm R=0$ whenever the pair form factor is nonzero, in agreement with Eq.~\eqref{eq:P}.
At co-directed massless configurations the amplitude vanishes; the amplitude-weighted shift
also vanishes, even though the phase of the amplitude is undefined.
The pair function \eqref{eq:R} coincides identically with the form \eqref{eq:P} of Appendix~\ref{sec:dirac},
obtained there from the Bloch-sphere embedding of the Weyl result \eqref{eq:weyl}; the
two are the same function written in the stereographic and in the Cartesian variables, and
either may be used. The numerical evaluation in Sec.~\ref{sec:hplot} uses the equivalent Cartesian expression
for $|V|^2\dr$, with its apparent pair denominators canceled analytically.

\section{Phase-space integrals for the generation profile}
\label{app:profile}

Writing the partner and transfer integrals of Eq.~\eqref{eq:G} over energies at fixed angles, one
encounters the following identities, where $\varpi$ is the collision energy transfer, distinct
from the external driving frequency $\omega$:
\begin{equation}
\begin{aligned}
&\int_{-\infty}^{\infty}\! d\xi_2\, f_0(\xi_2)\,[1-f_0(\xi_2-\varpi)] = \frac{\varpi}{e^{\varpi/T}-1}
= \varpi\,\nB(\varpi),\\
&\int_{-\infty}^{\infty}\! d\varpi\;\varpi\,\big[\nB(\varpi)+f_0(\varpi+\xi)\big]
= \frac{\xi^2+\pi^2T^2}{2},
\end{aligned}
\label{eq:FLintegrals}
\end{equation}
the second being the standard quasiparticle phase-space bracket [with $\varpi\to-\varpi$ the
first identity reads $\int d\xi_2\,f_0(\xi_2)[1-f_0(\xi_2+\varpi)] = \varpi[1+\nB(\varpi)]$].
These identities establish the overall $T^2$ phase space of the decay rate and, together with the
off-shell factor in the shift, the $T^3$ scaling of the generation term. They produce the familiar
$(\pi^2T^2+\xi^2)$ polynomial when the angular kernel is smooth. At a collinear endpoint the
cutoff depends on the integration energies, and the remaining dimensionless functions are those
in Eqs.~\eqref{eq:gprofile} and \eqref{eq:qprate}; no exact separation of energy and angle is
assumed.

\section{Low-temperature scaling and normalization of the Galerkin moments}
\label{app:sommerfeld}

The useful Sommerfeld moments are
\begin{equation}
\begin{aligned}
\Big\langle \xi^2\Big\rangle &\equiv
\int\!d\xi\,\xi^2\Big({-}\frac{\partial f_0}{\partial\xi}\Big)
=\frac{\pi^2T^2}{3},\\
\Big\langle \xi^4\Big\rangle&=\frac{7\pi^4T^4}{15},\qquad
\Big\langle \xi^2(\pi^2T^2+\xi^2)\Big\rangle=\frac{4\pi^4T^4}{5}.
\end{aligned}
\label{eq:moments2}
\end{equation}
The last identity explains the conventional $2\pi^4/5$ normalization below after the
first-harmonic angular average of one half. It is only a normalization convention: the exact
two-dimensional collision moments retain the nonseparable energy--angle dependence of the
collinear cutoffs.

\emph{(i) Clean drive moment.} Insert the projected trial function of
Eq.~\eqref{eq:chi-project} in Eq.~\eqref{eq:D-def}.  At leading Sommerfeld order both
$\bar\chi_1$ and $u_{\mathsf Q}$ reduce to the linear slope in Eq.~\eqref{eq:vperp}; with
$\nu(\varepsilon)\simeq\nu_F$ this gives
\begin{equation}
D[\chi] = \frac{\nu_F}{2}\frac{(1-m^2/\eF^2)^2}{p_F^2}\big\langle\xi^2\big\rangle
= \frac{\pi^2}{6}\nu_F T^2\frac{(1-m^2/\eF^2)^2}{p_F^2}.
\label{eq:Dresult}
\end{equation}

\emph{(ii) Generation moment.} Because the collinear cutoff correlates the angular and energy
integrations, we choose the following normalization and define the dimensionless generation
amplitude directly from the exact collision
moment \eqref{eq:J-def}:
\begin{equation}
\begin{aligned}
J[\chi]\equiv{}-\frac{2\pi^4}{5}\,
\mathcal A(m/\eF;T/\eF)(\nu_F\lambda)^2\nu_FT^4\frac{1-m^2/\eF^2}{\eF p_F^3v_F}.
\end{aligned}
\label{eq:Jresult}
\end{equation}
This definition includes both collinear endpoints and the full $\xi/T$ dependence of the kernel.
For $m>0$ in the sampled parameter range, the numerical evaluation gives $\mathcal A>0$, hence $J[\chi]<0$ with the choice
$\chi=u_{\mathsf Q}$. Together with $D[\chi],Q[\chi]>0$, this gives the positive shift conductivity of
Eq.~\eqref{eq:sigmaAB}.

\emph{(iii) Quadratic form.} With the same conventional normalization, the relaxation amplitude
is defined from the exact
entropy-production form \eqref{eq:Q-def}:
\begin{equation}
\begin{aligned}
Q[\chi]\equiv{}\frac{2\pi^4}{5}\,
\mathcal B(m/\eF;T/\eF)(\nu_F\lambda)^2\nu_FT^4\frac{(1-m^2/\eF^2)^2}{p_F^3v_F},
\end{aligned}
\label{eq:Qresult}
\end{equation}
with $\mathcal B>0$. It is essential that $\mathcal B$ be extracted from
$|\Delta(\chi\hat\pp)|^2$; replacing the collision operator by the quasiparticle loss rate
\eqref{eq:qprate} would omit gain terms and would violate the momentum zero mode. The numerical
evaluation reported in Sec.~\ref{sec:hplot} uses the full quadratic form.

\section{Exact momentum-mode elimination and thermal disorder moments}
\label{app:disorder-projection}

This appendix gives the finite-disorder operator underlying the leading hydrodynamic response.
In the notation of Eqs.~\eqref{eq:momentum-projectors}--\eqref{eq:projected-BE}, eliminating
$a_v$ gives $\hat S_\omega\psi_v=eE_xv_{{\rm eff},\omega}$, where
\begin{equation}
\hat S_\omega=\hat C_{\mathsf Q}+\mathsf Q\hat D\mathsf Q-i\omega
-\frac{|d\rangle\langle d|}{\mathcal N(\Gamma-i\omega)}.
\label{eq:exact-Schur}
\end{equation}
This is the Schur complement of the momentum block of the full Boltzmann operator.
The anomalous solution is
\begin{equation}
\psi_g=-eE_y\hat S_\omega^{-1}g_y,\qquad
 a_g=-\frac{\langle d|\psi_g\rangle}{\mathcal N(\Gamma-i\omega)}.
\label{eq:exact-anomalous}
\end{equation}
Combining its current with the equal accumulation contribution gives the exact result within
the additive collision model,
\begin{equation}
\sigma_{xy}^{\rm shift}(\omega)
=-2e^2\langle v_{{\rm eff},\omega}|\hat S_\omega^{-1}|g_y\rangle.
\label{eq:exact-reduced-Hall}
\end{equation}
All finite-frequency pairings are bilinear. This formula does not conjugate the frequency-dependent
effective velocity.

At dc, $\hat S_0=\hat C_{\mathsf Q}+\mathsf Q\hat D\mathsf Q-|d\rangle\langle d|/\mathcal D$.
For a real $\psi\in\mathsf Q$, the impurity correction is nonnegative:
\begin{equation}
\langle\psi|\hat D|\psi\rangle
-\frac{\langle\Phi_0|\hat D|\psi\rangle^2}{\mathcal D}\geq0.
\label{eq:Schur-positive}
\end{equation}
This follows from the Cauchy--Schwarz inequality in the $\hat D$-weighted metric.
The correction is $O(\Gamma)$; the inertial term is $O(\omega)$, and the rank-one term
is bounded on the real-frequency axis by its dc scale. Therefore replacing
$\hat S_\omega^{-1}$ by $\hat R_{\rm ee}$ on the relaxing subspace is controlled by
$(\Gamma+\omega)/\gamma_{\rm ee}$. It is this approximation,
not an expansion of $\omega/\Gamma$, that produces Eq.~\eqref{eq:frequency-interpolation}.

The effective dc drive needs only three one-particle moments. With
$w(\varepsilon)=-\partial_\varepsilon f_0$ and $\mu=\eF$, their exact conduction-band forms are
\begin{equation}
\begin{aligned}
\mathcal N&=\frac{1}{4\pi v^4}\int_{|m|}^{\infty}d\varepsilon\,w\varepsilon(\varepsilon^2-m^2),\\
\mathcal V&=\frac{1}{4\pi v^2}\int_{|m|}^{\infty}d\varepsilon\,w(\varepsilon^2-m^2),\\
\mathcal D&=\frac{\kappa}{4\pi v^4}\int_{|m|}^{\infty}d\varepsilon\,w
 (\varepsilon^2-m^2)(\varepsilon^2+3m^2).
\end{aligned}
\label{eq:disorder-moments-exact}
\end{equation}
Extending the thermal integrals below the band edge gives, with exponentially small errors
for $T\ll\eF-|m|$,
\begin{equation}
\begin{aligned}
\mathcal N&\simeq\frac{\eF(\eF^2-m^2)+\pi^2\eF T^2}{4\pi v^4},\\
\mathcal V&\simeq\frac{\eF^2-m^2+\pi^2T^2/3}{4\pi v^2},\\
\mathcal D&\simeq\frac{\kappa}{4\pi v^4}\Big[(\eF^2-m^2)(\eF^2+3m^2)\\
&\hspace{8mm}+2\pi^2T^2(\eF^2+m^2/3)+7\pi^4T^4/15\Big].
\end{aligned}
\label{eq:disorder-moments-Sommerfeld}
\end{equation}
The finite-$T$ comparison uses Eq.~\eqref{eq:disorder-moments-exact}, not an extrapolation of
these polynomial expressions near the band edge. The overall factor $\kappa$ cancels in
$\gamma_{\rm tr}(\varepsilon)\mathcal V/\mathcal D$, whereas
$\Gamma=\mathcal D/\mathcal N=\gamma_{\rm tr}(\eF)$ only to leading thermal order.

\bibliography{biblio-HAHE}

\end{document}